\documentclass[longauth]{aa}  

\bibpunct{(}{)}{;}{a}{}{,}
\usepackage{amsmath}
\usepackage{soul}

\usepackage{graphicx}
\usepackage{txfonts}
\usepackage{hyperref}
\usepackage{natbib}
\hypersetup{colorlinks=true,citecolor=blue}
\usepackage{multicol}
\usepackage{multirow}

\usepackage{caption}
\usepackage{subcaption}
\begin{document}

        \title{The VIMOS Public Extragalactic Redshift Survey (VIPERS)}
        
        \subtitle{Star formation history of passive red galaxies\thanks{based on
                        observations collected at the European Southern Observatory, Cerro Paranal, Chile, using the Very Large Telescope under programs 182.A-0886 and partly 070.A-9007.
                        Also based on observations obtained with MegaPrime/MegaCam, a joint project of CFHT and CEA/DAPNIA, at the Canada-France-Hawaii Telescope (CFHT), which is operated by the
                        National Research Council (NRC) of Canada, the Institut National des Sciences de l’Univers of the Centre National de la Recherche Scientifique (CNRS) of France, and the University of Hawaii. This work is based in part on data products produced at TERAPIX and the Canadian Astronomy Data Centre as part of the Canada-France-Hawaii Telescope Legacy Survey, a collaborative project of NRC and CNRS. The VIPERS web site is  \url{http://www.vipers.inaf.it/}.}}
        
        \author{M. Siudek\inst{1}\and
                K. Małek\inst{2}\and
                M. Scodeggio\inst{3}\and
                B. Garilli\inst{3}\and
                A.~Pollo\inst{2,4}\and 
                C.~P.~Haines\inst{5}\and
                A. Fritz\inst{3}\and
                M.~Bolzonella\inst{6}\and
                S.~de la Torre\inst{7}\and
                B.~R.~Granett\inst{5}\and
                L.~Guzzo\inst{5,8}\and
                U.~Abbas\inst{9}\and
                C.~Adami\inst{7}\and
                D.~Bottini\inst{3}\and
                A.~Cappi\inst{6,10}\and
                O.~Cucciati\inst{6}\and
                G.~De Lucia\inst{11}\and
                I.~Davidzon\inst{7,6}\and
                P.~Franzetti\inst{3}\and
                A.~Iovino\inst{5}\and
                J.~Krywult\inst{12}\and
                V.~Le Brun\inst{7}\and
                O.~Le F\`evre\inst{7}\and
                D.~Maccagni\inst{3}\and
                A.~Marchetti\inst{3}\and
                F.~Marulli\inst{13,6,14}\and
                M.~Polletta\inst{3,15}\and
                L.~A.~M.~Tasca\inst{7}\and
                R.~Tojeiro\inst{16}\and
                D.~Vergani\inst{17}\and
                A.~Zanichelli\inst{18}\and
                S.~Arnouts\inst{7}\and
                J.~Bel\inst{19}\and
                E.~Branchini\inst{20,21,22}\and
                O.~Ilbert\inst{7}\and
                A.~Gargiulo\inst{3}\and
                L.~Moscardini\inst{13,6,14}\and
                T.~T.~Takeuchi\inst{23}\and
                G.~Zamorani\inst{6}
        }
        
        \institute{
                Center for Theoretical Physics, Al. Lotnikow 32/46, 02-668 Warsaw, Poland; \\ 
                \email{gsiudek@cft.edu.pl}
                \and
                National Center for Nuclear Research, ul. A. Soltana 7, 05-400 Otwock, Poland; 
                \and
                INAF - Istituto di Astrofisica Spaziale e Fisica Cosmica Milano, via Bassini 15, 20133 Milano, Italy; 
                \and
                Astronomical Observatory of the Jagiellonian University, Orla 171, 30-001 Cracow, Poland; 
                \and
                INAF - Osservatorio Astronomico di Brera, Via Brera 28, 20122 Milano, via E. Bianchi 46, 23807 Merate, Italy; 
                \and
                INAF - Osservatorio Astronomico di Bologna, via Ranzani 1, I-40127, Bologna, Italy;
                \and
                Aix Marseille Universit\'e, CNRS, LAM (Laboratoire d'Astrophysique de Marseille) UMR 7326, 13388, Marseille, France;
                \and
                Dipartimento di Fisica, Universit\`a di Milano-Bicocca, P.zza della Scienza 3, I-20126 Milano, Italy; 
                \and
                INAF - Osservatorio Astronomico di Torino, 10025 Pino Torinese, Italy; 
                \and
                Laboratoire Lagrange, UMR7293, Universit\'e de Nice Sophia Antipolis, CNRS, Observatoire de la C\^ote d’Azur, 06300 Nice, France; 
                \and
                INAF - Osservatorio Astronomico di Trieste, via G. B. Tiepolo 11, 34143 Trieste, Italy; 
                \and
                Institute of Physics, Jan Kochanowski University, ul. Swietokrzyska 15, 25-406 Kielce, Poland; 
                \and
                Dipartimento di Fisica e Astronomia - Alma Mater Studiorum Universit\`{a} di Bologna, viale Berti Pichat 6/2, I-40127 Bologna, Italy; 
                \and
                INFN, Sezione di Bologna, viale Berti Pichat 6/2, I-40127 Bologna, Italy 
                \and
                IRAP,  9 av. du colonel Roche, BP 44346, F-31028 Toulouse cedex 4, France
                \and
                Institute of Cosmology and Gravitation, Dennis Sciama Building, University of Portsmouth, Burnaby Road, Portsmouth, PO1 3FX; 
                \and
                INAF - Istituto di Astrofisica Spaziale e Fisica Cosmica Bologna, via Gobetti 101, I-40129 Bologna, Italy; 
                \and
                INAF - Istituto di Radioastronomia, via Gobetti 101, I-40129, Bologna, Italy; 
                \and
                Aix Marseille Universit\'e, CNRS, CPT, UMR 7332, 13288 Marseille, France; 
                \and
                Dipartimento di Matematica e Fisica, Universit\`{a} degli Studi Roma Tre, via della Vasca Navale 84, 00146 Roma, Italy;  
                \and
                INFN, Sezione di Roma Tre, via della Vasca Navale 84, I-00146 Roma, Italy; 
                \and
                INAF - Osservatorio Astronomico di Roma, via Frascati 33, I-00040 Monte Porzio Catone (RM), Italy; 
                \and
                Division of Particle and Astrophysical Science, Nagoya University, Furo-cho, Chikusa-ku, 464-8602 Nagoya, Japan 
        }
        
        \date{Received September 15, 1996; accepted March 16, 1997}
        
        
        \abstract
        {}
        {We trace the evolution and the star formation history of passive red galaxies, using a subset of the VIMOS Public Extragalactic Redshift Survey (VIPERS). The detailed spectral analysis of stellar populations of intermediate-redshift passive red galaxies allows the build up of their stellar content to be
                followed over the last 8 billion years.}
        {We extracted a sample of passive red galaxies in the redshift range $0.4 < z < 1.0 $ and  stellar mass range 10 < $\rm{log(M_{star}/M_{\odot})}$ < 12 from the VIPERS survey. The sample was selected using an evolving cut in the rest-frame $U-V$ color distribution and additional cuts that ensured high quality. The spectra of passive red galaxies were stacked in narrow bins of stellar mass and redshift. We use the stacked spectra to measure the 4000 $\AA$ break ($D4000$) and the $H\delta$ Lick index ($H\delta_{A}$) with high precision. These spectral features are used as indicators of the star formation history of passive red galaxies. We compare the results with a grid of synthetic spectra to constrain the star formation epochs of these galaxies. We characterize the formation redshift-stellar mass relation for intermediate-redshift passive red galaxies.}
        {We find that at $z \sim 1$ stellar populations in low-mass passive red galaxies are younger than in high-mass passive red galaxies, similar to what is observed at the present epoch. Over the full analyzed redshift range $0.4 < z < 1.0 $ and stellar mass range 10 < $\rm{log(M_{star}/M_{\odot})}$ < 12, the $D4000$ index increases with redshift, while $H\delta_{A}$ gets lower. This implies that the  stellar populations are getting older with increasing stellar mass. Comparison to the spectra of passive red galaxies in the SDSS survey ($z\sim 0.2$) shows that the shape of the relations of $D4000$ and $H\delta_{A}$ with stellar mass has not changed significantly with redshift.
                Assuming a single burst formation, this implies that high-mass passive red galaxies formed their stars at $z_{form} \sim 1.7$, while low-mass galaxies formed their main stellar populations more recently, at $z_{form} \sim 1$. The consistency of these results, which were obtained using two independent estimators of the formation redshift ($D4000$ and $H\delta_{A}$), further strengthens a scenario in which star formation proceeds from higher to lower mass systems as time passes, i.e., what has become known as the downsizing picture. }
        {}
        
        \keywords{galaxies: formation --
                galaxies: evolution --
                galaxies: stellar content
        }
        \titlerunning{VIPERS: Star formation history of passive red galaxies}
        \maketitle
        %
        
        \section{Introduction}
        
        \label{sec:intro}

        According to the ~\cite{hubble1936} empirical tuning-fork diagram, elliptical (E) and lenticular (S0) galaxies form a group known as early-type galaxies (ETGs). Originally, the separation between ETGs and spiral galaxies was purely based on the lack of spiral arms in optical images~\citep{sandage}. However, with our improved understanding of galaxy properties, the early-type population can  be defined making use of a number of galaxy physical properties. Most commonly ETGs are described as red galaxies with old and passively evolving stellar populations and with no (or a negligible) sign of star formation. 
        In the local Universe, they are the most massive galaxies and host most of the stellar mass~\citep{baldry2}. These properties make them ideal laboratories to trace the history of stellar mass assembly and formation. Although the properties of ETGs have been extensively studied, leading to the discovery of a number of correlations among them, such as the so-called 
        fundamental plane, the color-magnitude relation, the Kormendy relation, or the Faber-Jackson relation~\citep[e.g.,][]{dressler,djorgovski,faber,kormendy,jeong,porter},        the physical process involved in their formation and evolution are still under debate. 
        
        Historically, two extreme scenarios for the star formation history (SFH) of ETGs have been proposed. 
        The classical monolithic collapse assumes that all parts of elliptical galaxies were formed at the same time after the 
        gravitational collapse of one or  more clumps of gas. According to this model the bulk of stars was formed 
        during a violent burst at high redshift, followed by a passive evolution~\citep{chiosi,romano,kaviraj}. 
        The alternative hierarchical model postulates that the galaxy formation is a continuous process and individual galaxies have 
        assembled their mass gradually through hierarchical merging of lower mass units and gas accretion, where most of the stellar 
        mass of the ETGs are assembled at relatively low redshift~\citep{davis1985,white,lucia}. As a consequence, the fact that massive ETGs 
        have been observed at relatively high redshift~\citep[e.g.,][]{papovich,Conselice,ilbert2013} has often been considered evidence of an anti-hierarchical 
        evolutionary scenario for these objects.
        
        In practice, over the last two decades, it has been shown that stellar mass plays a fundamental role in regulating star formation
        history (and therefore the evolution) of galaxies with low stellar mass systems having a star formation history significantly
        more extended in time than their high stellar mass counterparts~\cite[e.g.,][]{cowie,gavazzi1996,thomas2002,kauffmana}.
        As a result of this mass dependent evolution, generally known as downsizing~\citep{cowie}, stars were formed earlier and faster 
        in massive galaxies, and therefore these systems completed their star formation at higher redshifts than lower mass galaxies. 
        Furthermore one needs to consider the possibility that star formation history and mass assembly history do not necessarily
        proceed along the same path in a hierarchical scenario. Therefore it becomes necessary to extend the downsizing concept
        to stellar mass assembly; i.e., more massive galaxies have assembled their stellar masses before less massive galaxies~\cite[this is called mass-downsizing;][]{cimattid}. Evidence in favor of the downsizing scenario has been presented by a number of 
        authors~\citep[e.g.,][]{renzini, pozetti,cimattib,cimattia,  morescoage, fritzaa,fritz}. The most recent studies suggest that 
        the most massive ETGs (with stellar masses $\rm{>10^{11}}$ $\rm{M_{\odot}}$) have assembled their masses at relatively high redshifts 
        and that most of these were already in place since at least $z \sim 1$ (e.g.,~\citealp{pozetti}). However, 
        \cite{thomas2005,thomas2010} have shown that the most massive galaxies in the local Universe 
        ($\rm{log(M_{star}/M_{\odot}) \sim 12}$) were formed at redshift 3 - 5, in contrast with less massive galaxies 
        ($\rm{log(M_{star}/M_{\odot}) \sim 10.5}$) that formed at $z \sim 1$.

        One of the most direct methods to study the evolution of galaxies and their SFH is based on the measurements of the 4000$\AA{}$ 
        spectral break (hereafter $D4000$) and of the Balmer absorption line index $H\delta_{A}$.  These two spectroscopic indices are 
        commonly used as age indicators, not only for ETGs, but for the totality of the galaxy 
        population~\cite[e.g.,][]{hamilton, balogha, bruzual, kauffmana, moresco}.
        The $4000\AA$ break is the strongest discontinuity in the optical spectrum of a galaxy and is caused by the accumulation of 
        a large number of absorption lines in a narrow wavelength range,  mainly including ionized metals. 
        In young, hot stars the elements are multiply ionized and the line opacities decreases. This is reflected in the strength of $D4000$, 
        which is small for young stellar populations and becomes larger for older galaxies. It means that the break that  occurs at 
        4000$\AA$ is correlated with the age of the stellar population.  However, the age effects are often confused with the impact of  metallicity, as the increase of metallicity  may also result in stronger absorption features. 
        A galaxy becomes redder, as its grows old and more stars move to the giant branch, but also as  metallicity increases since the stellar photosphere become less transparent,  which results in cooling of the stars.  
        The inability to unambiguously separate the two effects (age and metallicity), well known as the age-metallicity degeneracy~\citep{worthey1994}, makes the determination of the age of the stellar population complicated. Different color indices are degenerated for old stellar populations (above age of 5~Gyr), while an increase/decrease of the age of population by a factor of three affects indices in a way that is practically identical with an increase/decrease in metallicity by a factor of two. This effect is commonly known as 3/2 rule~\citep{worthey1994}. The $D4000$ feature is also not fully discriminatory in the separation of age and metallicity effects. The mean metallicity sensitivity parameter, the age change, which is needed to balance the metallicity change so that the index remains constant, is on the level of 1.3~\citep[with larger numbers indicating greater metallicity sensitivity,][]{worthey1994} This degeneracy may make dating of old stellar populations unreliable. Thus a realistic spread of metallicities or at least high order Balmer lines (like $H\delta$ or $H\gamma$) should be included in the models for better age and metal discrimination.

        The $H\delta$ absorption line can also be used to date stellar populations. The line is hidden in galaxies with an ongoing 
        star formation because of the dominance of hot O and B stars, which have weak intrinsic absorption lines~\citep{balogha},
        in the galaxy spectrum, and the filling of the absorption feature by the emission coming from HII regions. After the star formation
        activity stops, the equivalent width of $H\delta$ absorption line strongly increases, reaching a maximum (almost 10$\AA$) 
        when the dominant stellar population in a galaxy consists mainly of the type A stars (approximately 1 Gyr after the termination of 
        star formation activity), and then decreases continuously as the stellar population ages even further~\citep{borgne}. Also in this case, 
        metallicity plays an important role to determine the strength of the line, as the mean metallicity sensitivity parameter for $H\delta$ line is 0.8--1.1~\citep{worthey}. The $H\delta$ is more suitable as an age indicator then $D4000$, as  its age sensitivity is higher. This makes the $H\delta$ line a very useful tool 
        to study the age of stellar populations and to analyze the SFH of any type of galaxies.

        Stellar ages determined on the basis of spectral features may also  suffer  from dust effects~\citep{worthey1994}. However, ETGs have little dust and, therefore, we do not expect their $D4000$ and $H\delta_{A}$ to be affected by dust attenuation~\cite[e.g.,][]{macarthur}.
        
        Used together measurements of $D4000$ and the $H\delta$ line allow for a significantly improved separation of galaxies into their 
        two main families with respect to a simple morphological
        classification: those with an ongoing star formation and those without an ongoing star formation. Studies of the conditional density distributions of $D4000$ and $H\delta$ features as a function of stellar mass 
        have shown that local Universe galaxies have a well-defined transition mass of $\rm{3 \times 10^{10} M_{\odot}}$ above 
        which red galaxies with high surface mass densities contain mainly an old stellar population, which dominates the overall galaxy 
        population. However, below this mass blue, low surface mass densities galaxies with ongoing star formation, dominate the 
        scene~\citep{kauffmana}. Earlier results based on galaxy color and morphology had shown indications of a similar 
        partitioning~\cite[e.g.,][]{scodeggio}, but could not identify such a clear transition between the two populations.
        
        In this paper we extend the analysis of~\cite{kauffmana} by presenting a study of the star formation history for a large sample 
        of red passive galaxies (ETGs selected based on the colors) with redshift ranging from 0.4 to 1.0 and stellar masses from $10^{10}$ to $10^{12}$ $\rm{M_{\odot}}$. 
        Our analysis is based on the two spectral features ($D4000$ and $H\delta$) measured on the composite spectra of passive red galaxies extracted from the VIPERS spectroscopic dataset~\citep{guzzo}. We compare the spectroscopic properties of the stacked spectra with a 
        grid of synthetic models to constrain the redshift of formation for these objects. To check the global evolution of passive red galaxies 
        from $z=0.1$ to $z=1.0,$ we compare the results with those obtained for passive red galaxies selected from the SDSS survey. 
        In terms of volume and sampling VIPERS can be considered the $z \sim 1$ equivalent of current state-of-the-art local ($z<0.2$) surveys such as the 2dFGRS~\citep{colles} 
        and Sloan Digital Sky Survey~\citep[SDSS; ][]{abazajian,york}. Thus, we are able to trace the SFH of passive red galaxies with a 
        comparable statistical significance.  
        
        The paper is organized in the following way. In Sect. \ref{sec:data} we present the VIPERS data sample and describe the procedure 
        for the selection of the passive population. In Sect.~\ref{sec:methedology} we introduce the two spectral indicators, 4000$\AA$ break 
        and $H\delta$ line, and describe the methodology for co-adding spectra. 
        In Sect. \ref{sec:results} we present our results of the analysis of the spectral indicators. We show the trends of these features in different redshift and stellar mass ranges for the passive red galaxies sample and compare these trends to the spectral properties of local passive red galaxies. We independently estimate the redshift of formation based on $D4000$ and $H\delta$ line indices by comparing them with a grid 
        of synthetic spectra. In Sect~\ref{sec:summary} we present a summary of our analysis. In App.~\ref{app:A} we show how the selection criterion may affect our results. 
        
        In all the presented analysis, magnitudes and colors are given in the Vega system unless specified otherwise. Our cosmological framework assumes $\Omega_{m}$ = 0.30, $\Omega_{\Lambda}$ = 0.70, and $h_{70}$ = $H_{0}/(70$ $km s^{-1} Mpc^{-1})$.
        
        \section{Data and sample description}\label{sec:data}
        
        \subsection{VIPERS}
        
        Our work is based on the galaxy sample from the  VIMOS Public Extragalactic Redshift Survey~\cite[VIPERS,][]{guzzo}. 
        This spectroscopic survey was designed to map in detail the large-scale spatial distribution and properties of galaxies 
        over an unprecedented volume of $\rm{5 \times 10^7 h^{-3}Mpc^3}$ at $0.5 < z < 1.5$. In total almost 100,000 galaxies were 
        observed to provide a relatively high sampling rate~\citep[$\sim 40 \%$,][]{guzzo} of the underlying galaxy population, which is 
        composed of galaxies with $\rm{i'_{AB}}<22.5$ over an area of $\rm{\sim 24~deg^{2}}$ and is contained within fields W1 and W4 of the 
        Canada-France-Hawaii Telescope Legacy Survey (CFHTLS)\footnote{ \url{http://www.cfht.hawaii.edu/Science/CFHTLS/}}. 
        A simple and robust pre-selection in the $(u-g)$  versus $(r-i)$ color-color plane was used to efficiently remove 
        galaxies with $ z < 0.5$~\citep{guzzo,garilli} from the parent photometric sample. Spectroscopic observations were carried out with 
        the VIsible Multi-Object Spectrograph~\cite[VIMOS;][]{fevre} mounted on the ESO Very Large Telescope, using the  
        multi-object spectroscopy (MOS) mode with the low-resolution red grism ($\rm{\lambda_{blaze} = 5810\AA{},}$ $\rm{R = 230, 1''}$ slit) 
        yielding a spectral coverage between 5500 and 9500$\AA{}$ with an internal dispersion of $\rm{7.14}\AA{}$ $\rm{pixel^{-1}}$~\citep{marco}. 
        The detailed survey description can be found in~\cite{guzzo}. The first data release (hereafter PDR1\footnote{\url{http://vipers.inaf.it/rel-pdr1.html}}) has been already published and detailed information about 
        the VIPERS sample can be found in~\cite{garilli}. The VIPERS database system and data reduction pipeline are 
        described in \cite{bianca}.
        
        The analysis presented in this paper is based on the VIPERS internal data release version 5, which contains spectroscopic measurements for 76,045 objects from the W1 and W4 fields, which corresponds to $85\%$ of the final sample.

        \subsection{Luminosities and stellar masses of VIPERS galaxies}
        
        Stellar masses for the VIPERS sample were estimated via spectral energy distribution 
        fitting~\citep[hereafter SEDs;][]{davidzon,davidzon2015} using the \textit{Hyperzmass} code~\citep{bolzonella}. 
        The program was used to fit model SEDs to the multi-band photometry (in the filters $u^{*}, g^{'}, r^{'}, i^{'}$, $z^{'}, K_s$) 
        and to select the best fitting template on the basis of the lowest derived $\chi^2$ value. A grid of SED models was built on the basis 
        of stellar population synthesis models from~\citealp{bruzual} (hereafter BC03), adopting the Chabrier initial mass 
        function~\citep{chabrier} with nonevolving stellar metallicity and with either constant or exponentially declining 
        star formation histories. In order to model the galaxy dust content, both the \cite{calzetti} and Pr\'{e}vot-Bouchet~\citep{prevot,bouchet} extinction laws, with extinction magnitudes ranging from 0 (corresponding to no dust) up to 3, were used. For a more detailed 
        description of the VIPERS SED fitting procedure we refer to~\cite{davidzon,davidzon2015}.
        Absolute magnitudes were computed starting from the apparent magnitude in the photometric filter that most closely matches the selected 
        rest-frame band to which a \textit{k }correction was applied based on the best-fitting model SED~\cite[details in][] {davidzon,fritz}.

        \subsection{Selection of passive red galaxies}\label{sec:etg}

        A strong bimodality in many galaxy properties, including colors (e.g.,~\citealp{bell},~\citealp{balogh2004},~\citealp{franzetti}), H$\alpha$~\citep{balogh}, $[OII]\lambda3727$ emission~\citep{mingoli}, 4000$\AA{}$ break~\citep{kauffmana}, or star formation history~\citep{brinchmann} has been observed at least up to $z \sim 1.5$. These observed bimodalities allow for a relatively
        simple separation of galaxies into red- and blue-type populations. However, as shown for example by~\cite{renzini}, ETG samples selected according to different criteria (photometry, morphology, and spectroscopy) do not fully overlap. For example, out of the spectroscopically selected SDSS ETGs, only 55$\%$ satisfy the equivalent morphological selection, and 70$\%$ the equivalent color criterion~\citep[][Table 1]{renzini}. 
        As a result, samples of ETGs selected by different methods are roughly similar, but not all the same~\citep{morescocolor}. 
        The most important reason behind these differences is the fact that most ETG samples are affected by some level of contamination that 
        primarily comes from  dust-reddened galaxies with relatively low levels of star formation activity, and this contamination may strongly 
        affect the derived mean properties of the ETGs population. It has been shown that a contamination at the level of a few percent 
        from young stars can drastically alter the colors and spectra of a passive population even if the majority of its mass is provided by 
        old stars~\citep{trager}. Therefore, to derive meaningful mean properties for the population of ETGs it is essential 
        to build a  sample with contamination as low as possible, even at the cost of a reduced sample completeness. 
        
        In case of our analysis we are focused on the ETGs selected based on colors rather than morphology, and therefore we use nomenclature red, blue, and green galaxies. General criteria to separate the red- and blue-type galaxy populations within the VIPERS dataset were already discussed by~\cite{fritz}. 
        In this work we focus on a sample of passive red galaxies with the lowest possible contamination and with the best possible 
        resulting completeness (better than 75\%), up to $z = 1$. Thus, we base our passive red galaxies selection on the so-called bimodality 
        criterion of~\cite{fritz}. In that work it was shown, based on a comparison between a full spectrophotometric classification and a simple 
        color selection criterion, that it is possible to use a cut in galaxy $U-V$ color that evolves with redshift~\cite[e.g.,][]{wolf, peng} 
        to obtain a sample of red passive galaxies with almost constant and high ($\sim85\%$) completeness up to $z = 1$ and a contamination from intermediate 
        (green valley) and blue galaxies that remains lower than $10\%$ until $z = 0.8$ and then reaches $\rm{\sim 30 \%}$ at higher redshifts. 
        This bimodality criterion differs from similar criteria used in other works. It separates galaxies into three classes (red, blue, and green types) with the goal to reduce the contamination by  galaxies that have not yet reached the purely passive evolution stage of red passive galaxies. 
        As a result of this choice, the $U-V$ color cut used to isolate 
        red passive galaxies does not coincide with the value at the position of the local minimum in the global galaxy color distribution. 
        Of course, since different populations overlap in the boundary region,
        it is possible that this specific color cut 
        could lead to some level of sample incompleteness, which would manifest itself as a loss of a fraction of the bluest objects among 
        the true red passive galaxies.  A quantitative estimate of the effects that this possible incompleteness could have on our results is presented in Appendix~\ref{app:A}.
        
        Our initial sample consists of VIPERS galaxies from the W1 and W4 fields covering the redshift range $0.4 < z < 1.0$, and with VIPERS redshift measurement confidence flag $(z_{flg})$ 3 
        and 4. As shown by~\cite{guzzo}, flags 3 and 4 are not separable and indicate a reliable redshift measurement corresponding to a confidence in the redshift measurement at the level of $99.6\%$. This confidence flag selection is effectively a selection in spectra signal-to-noise (S/N) ratio and is dictated by the spectral
        features measurement criteria on stacked spectra discussed in Sect.~\ref{sec:stacking}. Given the  wavelength range used for spectra 
        normalization ($4010-4600\AA$ in the rest frame; see Sect.~\ref{sec:stacking}) and the spectral coverage of the VIPERS survey 
        ($5500-9500\AA$), we decided to limit our sample to galaxies with redshift $z<1$ to measure both spectral features of interest 
        ($H\delta$ and $D4000$) with the same quality.  The variable $U-V$ color cut from~\cite{fritz} is given by the relation 
        $\rm{(U-V) = 1.1 - 0.25\times}$ z (color in Vega system), and results in a sample of 8,174 candidate passive red galaxies.
        
        This sample is further pruned to reduce residual contamination from star-forming galaxies and to eliminate spectra that are not
        suitable for the stacking procedure. In particular we removed from our sample those spectra that are affected by stronger than average
        noise in the rest-frame wavelength range, in which $D4000$ and $H\delta$ are measured ($\rm{3850-4250\AA{}}$), because of either fringing, strong 
        sky subtraction residuals, or the presence of a zeroth-order spectrum from a bright object located in an adjacent MOS slit. 
        We eliminated 3,295 galaxies affected by such defects ($40.3\%$ of the passive galaxy sample).
        Moreover, we performed a visual inspection of all remaining spectra, and rejected all those showing some evidence of ongoing star 
        formation via the presence of the H$\delta$ and the $[OII]\lambda3727$ line in emission. This final pruning has lead to the 
        rejection of additional 888 spectra from our sample ($10.9\%$ of the passive galaxy sample).

\begin{figure*}[]
                \centering
                \centering
                \includegraphics[width=0.95\textwidth]{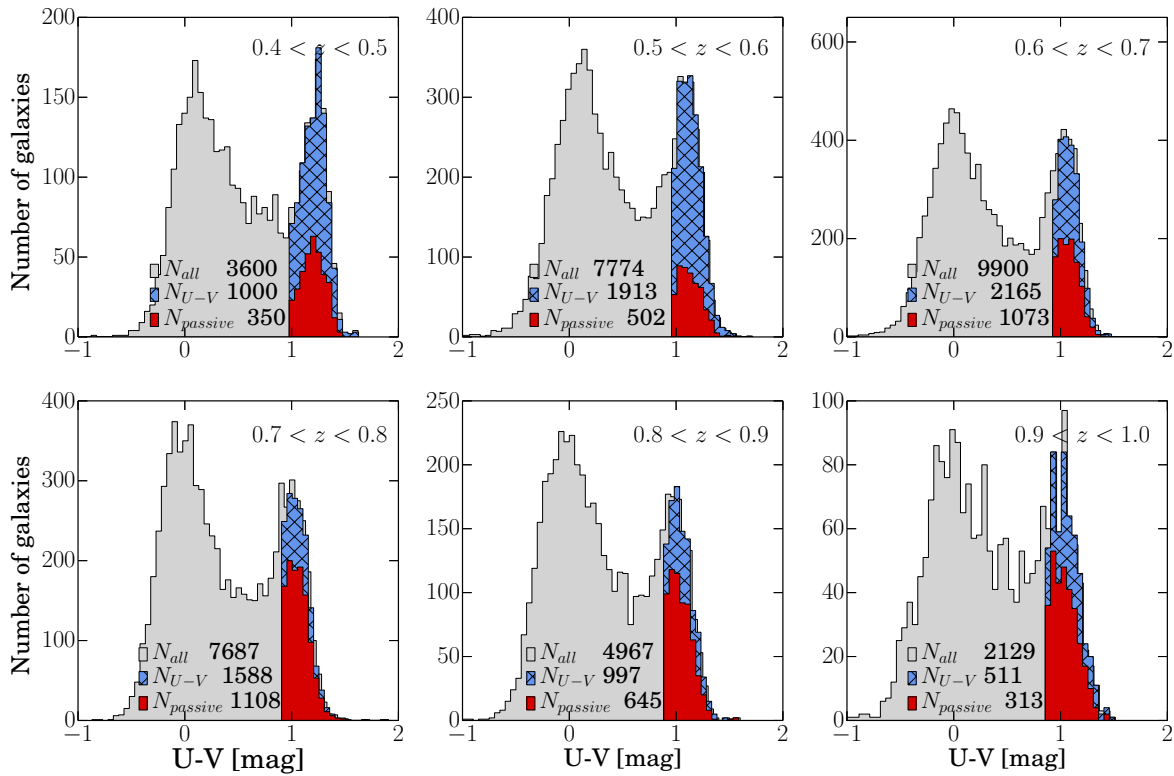}
                \caption{Distribution of rest-frame $U-V$ color of the initial sample of VIPERS galaxies (in gray), the passive galaxy sample defined solely on the basis of $U-V$ color (in blue), and the final ETG sample (in red) in different redshift bins in the range of $0.4 < z < 1.0$ as labeled in each panel. The  separation of red, passive galaxies from blue late-type galaxies was defined by an evolving cut in the $U-V$ color distribution~\citep{fritz}. Colors are given in the Vega system.}
                \label{fig:bimodaluv}
        \end{figure*}

        The resulting final sample of passive red galaxies with high quality spectra is composed of 3,991 galaxies. 
        Fig.~\ref{fig:bimodaluv} shows, for different redshift bins, the rest-frame $U-V$ color distributions for the full VIPERS sample 
        (in gray) for the sample of candidate passive red galaxies obtained using the $U-V$ color cut (in dashed blue) and for final 
        sample of passive objects with high quality spectra (in red). 
        
        One further independent check on the residual presence of dusty red spirals inside our passive galaxy sample was carried out 
        using the Sersic index derived for VIPERS galaxies~\citep{krywult} using r-band CFHTLS images. We estimate the possible contamination 
        by dusty spiral galaxies, identified as those with measured Sersic index $n < 1.5$, to be less than $6\%$ and, therefore, we decided against any
        further pruning of our passive galaxy sample. The choice of a relatively low value for the Sersic index boundary is justified by the fact
        that the index measurements, carried out on ground-based images, are biased toward lower values than typically measured for fully resolved
        galaxies in the local Universe or for high-redshift objects observed with the Hubble Space Telescope (HST). 
        
        Finally, we examined the possibility that the VIPERS sample definition, based on cuts in the g, r, i color-color 
        plane~\citep[~$(r-i)~>~0.5\cdot(u-g)$~or $(r-i)>0.7$,][]{guzzo}, could bias our passive galaxy sample over the redshift 
        range $0.4 < z < 0.6$, where the completeness of the full VIPERS sample gradually changes from zero to 100\%. As it turns out this
        color-color cut, which is removing bluer than average galaxies in that redshift range, is not affecting our passive galaxy sample. 
        In fact we observe that the distribution of the $D4000$ break in individual passive galaxy spectra is independent from the distance that 
        these galaxies have from the above color-color cut, and therefore we conclude that our 
        passive galaxy sample is not significantly biased toward extremely red (and therefore old) objects even in the redshift range $0.4 < z < 0.6$ .

        \section{Methodology}\label{sec:methedology}
        
        \subsection{Spectral indicators}\label{sec:sp}
        
\begin{figure}[t]
                \centering
                \includegraphics[width=0.49\textwidth]{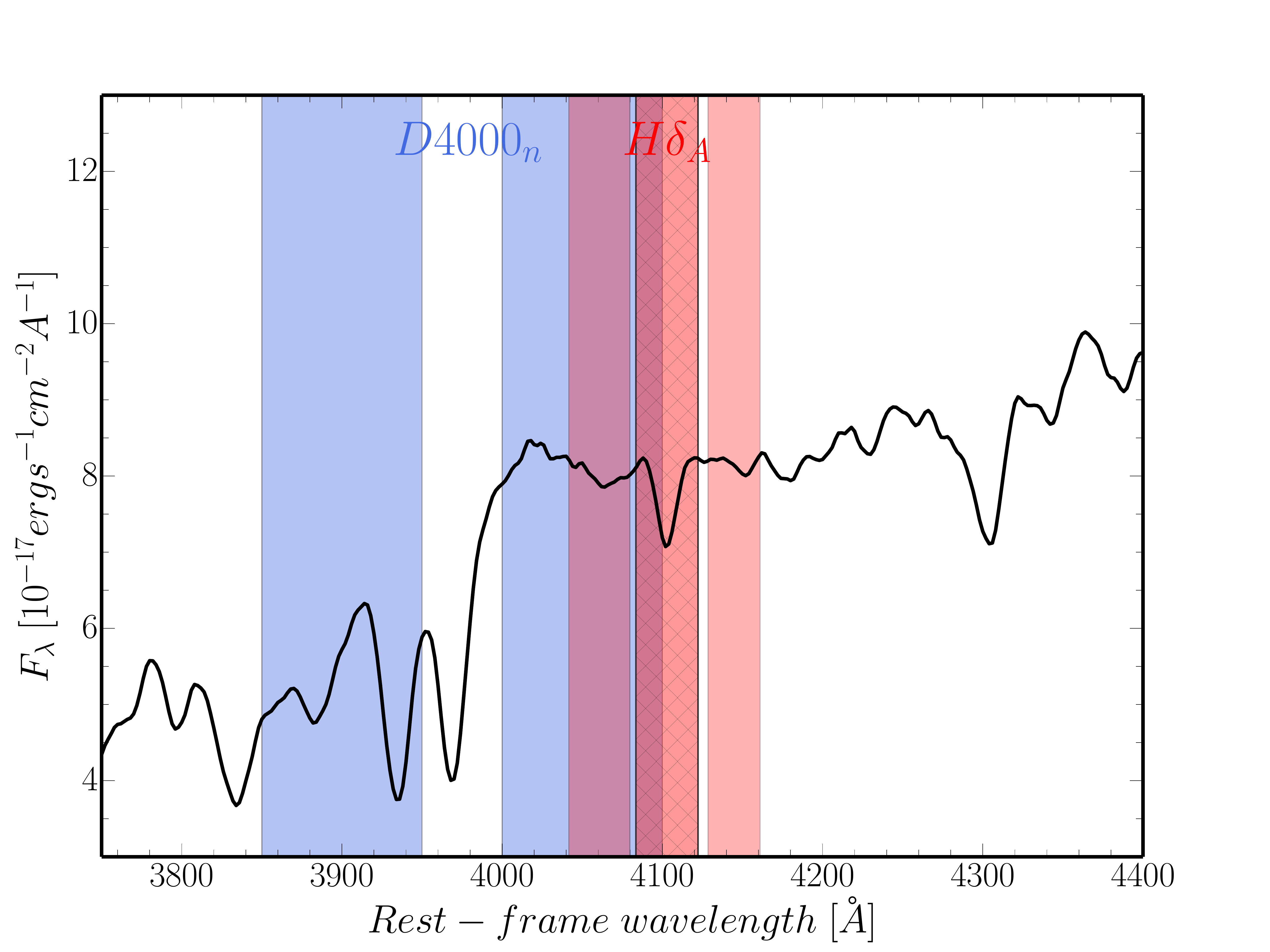}
                \caption{Exemplary stacked spectrum of passive red galaxies taken from the VIPERS database in the wavelength range 3800 - 5000$\AA{}$. Blue shaded areas show the ranges used to evaluate the $D4000_{n}$ break. Red regions  correspond to pseudocontinua for the $H\delta_{A}$, while the hatched area shows the $H\delta_{A}$  bandpass.  }
                \label{fig:stacked}
        \end{figure}

        In this work we use the $D4000$ and $H\delta$ indicators as tools to reconstruct the star formation history of passive red galaxies. We adopt the narrow definition of the $D4000$ spectral indicator that is  presented in~\cite{balogha} because it is less sensitive to the reddening effect in comparison to the definition presented by~\cite{bruzual83}. We denote this index as $D4000_{n}$ and define it as the ratio between the continuum flux densities in a red band ($4000-4100\AA$) and a blue band ($3850-3950\AA$),
        
        \begin{equation}
                D4000_{n} = \frac
                {(\lambda_{2}^{blue}-\lambda_{1}^{blue})
                        \int 
                        \limits_{\lambda_{1}^{red}}^{\lambda_{2}^{red}}
                        {F_{\nu}d\lambda}
                }
                {(\lambda_{2}^{red}-\lambda_{1}^{red})
                        \int 
                        \limits_{\lambda_{1}^{blue}}^{\lambda_{2}^{blue}}
                        {F_{\nu}d\lambda}
                },
        \end{equation}
        The spectral regions used to calculate $D4000_{n}$ are indicated  in blue in Fig.~\ref{fig:stacked}.
        
        For the  $H\delta$ line we use the $H\delta$ Lick index (hereafter $H\delta_{A}$), one among a set of 21 spectral indices 
        known as the Lick-IDS system~\citep{wortheya}, defined by~\cite{worthey} as
        
        \begin{equation}
                H\delta_{A} = (\lambda_{2} - \lambda_{1})\cdot (1-{F_{I}}/{F_{C}}), 
        \end{equation}
        where $F_{I}$ is defined as the continuum flux minus the absorption and $F_{C}$ is the continuum flux; $\lambda_{2}-\lambda_{1}$ corresponds to the width of the bandpass used to measure the index. The absorption line strength is obtained by comparing measurements of the spectral
        flux in the central feature bandpass and in two flanking pseudocontinuum regions. For the $H\delta_{A}$ index the feature range is 4083.50 - 4122.25$\AA{}$, the blue continuum range is 4041.60 - 4079.75$\AA{}$, and the red continuum range is 4128.50 - 4161.00$\AA{}$. The spectral regions used to calculate this index are indicated  in red in Fig.~\ref{fig:stacked}.
        
        \subsection{Stacking procedure}\label{sec:stacking}
        
        The typical S/N ratio  of VIPERS spectra is enough to measure the strength of the 4000$\AA$ break 
        in individual spectra accurately, but it is not sufficient to detect or measure the $H\delta$ line with sufficient accuracy.  Thus, we co-added 
        individual galaxy spectra to construct a high S/N ratio composite spectrum, as carried out in previous studies on the evolution of stellar populations in galaxies over the redshift range $0 < z < 1.5$~\cite[e.g.,][]{schiavon,sanchez, onodera, onodera2015,andrews, morescocolor,choi,price}. 
        
\begin{figure}[]
                \centering
                \centering
                \includegraphics[width=0.49\textwidth]{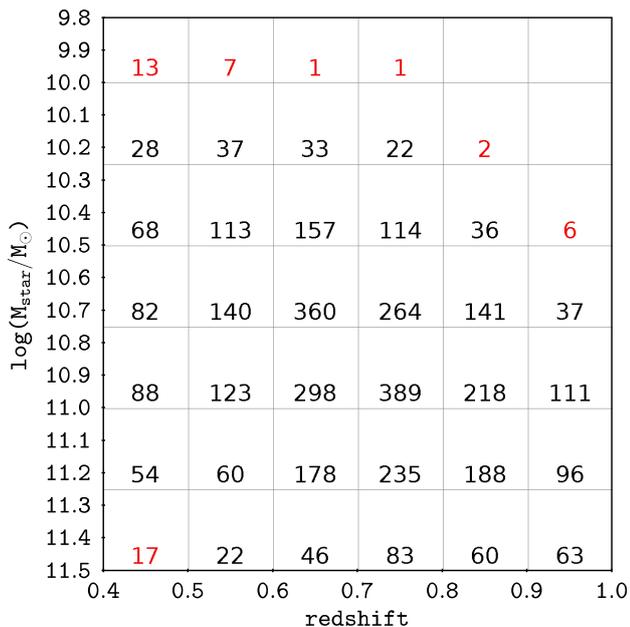}
                \caption{Final numbers of VIPERS spectra of passive red galaxies for different redshift and stellar mass bins. Bins with insufficient number of galaxies ($<20$), shown in red, are not used in the following analysis.} 
                \label{fig:number}
        \end{figure}

        \cite{fabello} have found that stacking spectra  provides the means to measure spectral features up to one order of magnitude below the detection limit for individual spectra, before non-Gaussian noise becomes dominant, or even deeper if the noise is well characterized~\citep{delhaize}.  Co-adding the rest-frame spectra of $N$ galaxies should reduce the root mean square (rms) noise of the resulting spectrum as $1/\sqrt{N}$ up to $N\sim 300$. When stacking more galaxies, the non-Gaussian noise dominates and the rms of the resulting spectrum is still reduced, but at a lower rate~\citep{fabello}. Thus, if we co-add a sufficient number of galaxies, the 
        $H\delta$ line becomes measurable in the final stacked spectra, even if it is indistinguishable from the noise 
        in the single object observations.
        Given the relatively large sample of VIPERS passive red galaxies we have available (3,991 objects), we are able to stack spectra in narrow redshift and stellar mass bins, partitioning our dataset in the following way:
        \begin{itemize}
                \item{six redshifts bins ($\delta$z = 0.1 from 0.4 to 1.0),}
                \item{six stellar mass bins ($\delta$$\rm{M_{star}}$ = 0.25 dex over the range of $10.00 < \rm{log(M_{star}/M_{\odot}) < 11.25}$ and a wider bin in the range of $11.25 < \rm{log(M_{star}/M_{\odot}) < 12.00}$ because of the 
                        relative scarcity of high-mass passive red galaxies. The lower stellar mass limit is set to $\rm{log(M_{star}/M_{\odot}) = 10.00}$ owing to the overall mass limit of the spectroscopic VIPERS sample.}
        \end{itemize}
        A table with the number of passive red galaxies in each stellar mass and redshift bin is shown in Fig.~\ref{fig:number}.

        The stacking procedure we applied to VIPERS passive galaxy spectra can be described as follows:    (1) individual spectra are shifted to  rest frame and resampled to a constant dispersion and a common wavelength grid, (2) a scaling factor is derived for each spectrum using
        the median flux computed  in a wavelength region between 4010 and 4600$\AA$, (3) individual rest-frame spectra are normalized by dividing
        the flux at all wavelengths by the above scaling factor. With this scaling we preserve the equivalent width of lines, but not their 
        total flux; (4) the stacked spectrum is obtained by computing the mean flux from all individual spectra at all wavelengths in the common
        wavelength grid, and (5) the final stacked spectrum is rescaled by multiplying the flux at all wavelengths by the average value of the individual spectra scaling factors.

\begin{figure}[]
                \centering
                \centering
                \includegraphics[width=0.49\textwidth]{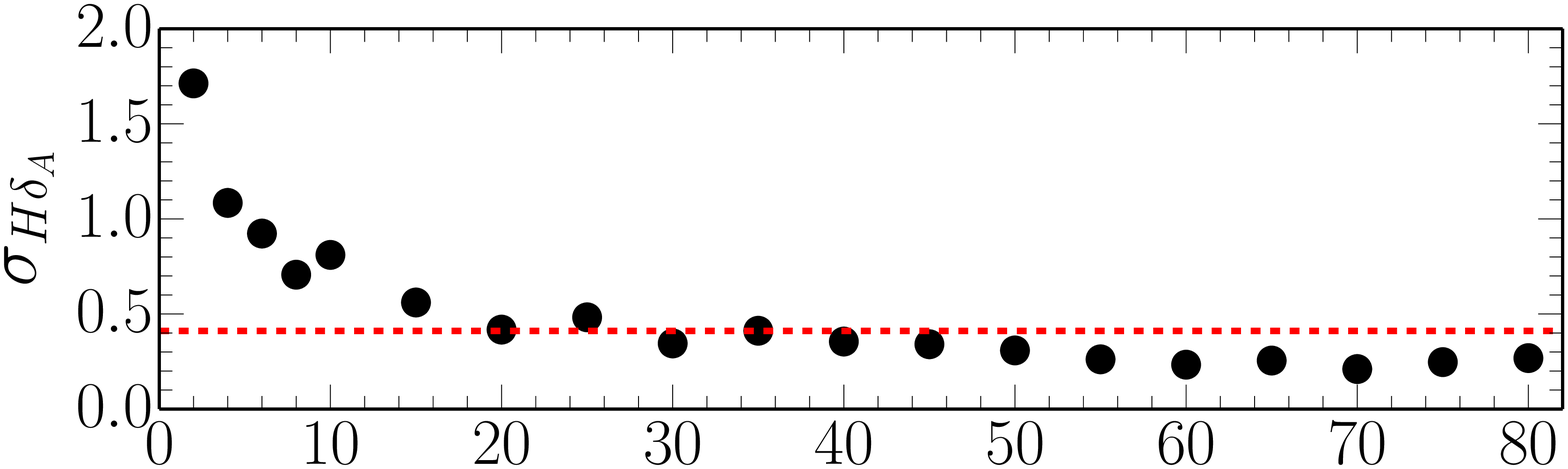}
                \includegraphics[width=0.49\textwidth]{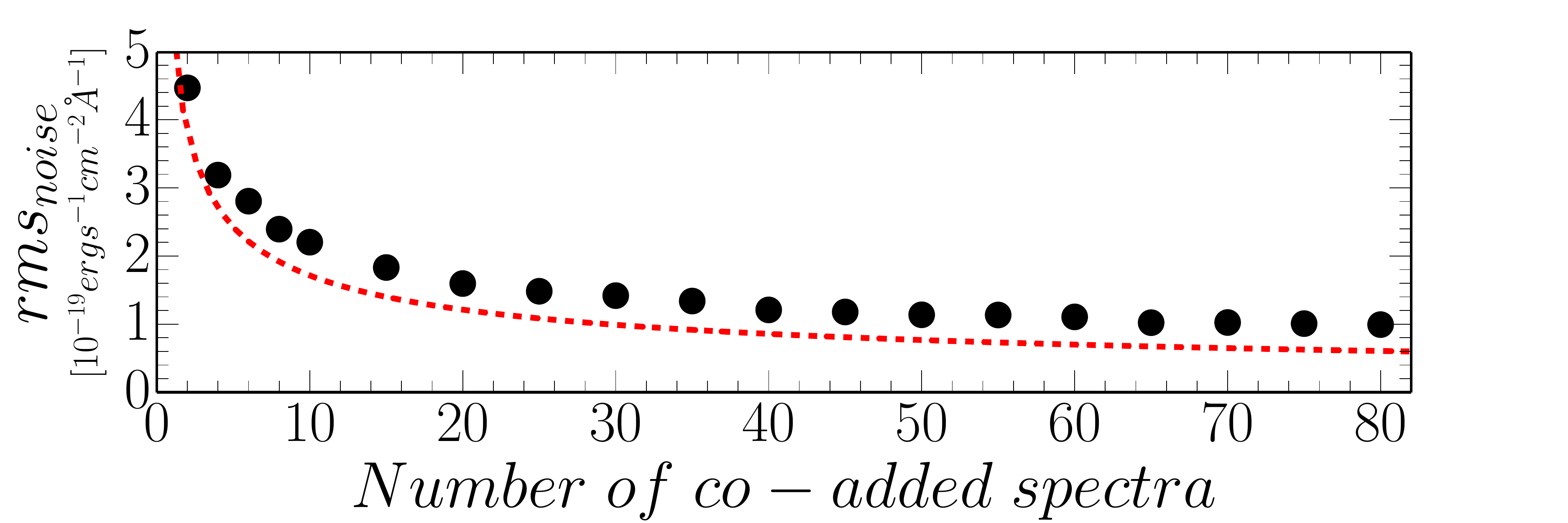}
                \caption{       
                        Distribution of standard deviations for $H\delta_{A}$ (upper panel) and of rms of noise around $H\delta$ line as a function of the number of co-added spectra. The red dashed line  represents 3  times the standard deviation calculated for spectra composed of all 360 galaxies (upper panel) and median rms of noise reduced by $1/\sqrt{N}$ (lower panel).   
                }
                \label{fig:minnumber}
        \end{figure}

        To quantify the minimum number of spectra that needs to be stacked to provide a robust mean spectrum (in terms of measurements of the
        $H\delta_{A}$ index) within a given stellar mass and redshift bin, we performed a test using the set of 360 galaxies in the redshift range $0.6 < z < 0.7$ and stellar mass in the range 10.5 < $\rm{log(M_{star}/M_{\odot})}$ < 10.75. We divided this set 
        into independent subsets, which are all composed of a given number of galaxies, and obtained a stacked spectrum for each subset. We then measured 
        the $H\delta_{A}$ index and the noise in the spectral continuum around the $H\delta$ line on each stacked spectrum. Finally, we computed 
        the mean value and standard deviation of all these measurements and repeated the same exercise for different sizes of the independent 
        subsets; the number of galaxies in each subset could be 2, 4, 6, 8, 10, 15, 20, 25, 30, 35, 40, 45, 50, 55, 60, 65, 70, 75, and 80, 
        resulting into 180, 90  60, 45, 36, 24, 18, 14, 12, 10, 9,  8, 7, 6, 5, 5, 4, and  4 independent stacked spectra, respectively. 
        The distributions of the standard deviation in the $H\delta_{A}$ index measurements, and of the average continuum rms noise around the $H\delta$ line as a function of subset size are shown in Fig.~\ref{fig:minnumber}. The red dashed line   in the upper panel represents three times the uncertainty in the $H\delta_{A}$ index measurement calculated on the stacked spectrum obtained
        using all the 360 galaxies, whereas in the lower panel it represents the expected reduction of the spectral continuum noise in the stacked spectra by a factor of $1/\sqrt{N}$.  
        Based on the results of this test we conclude that to obtain a stacked spectrum that allows a reliable measurement of spectral features, 
        which are representative of the true mean properties of passive red galaxies in a given stellar mass and redshift bin, we need to co-add at 
        least 20 individual spectra. Bins with insufficient number of galaxies are indicated in red in Fig.~\ref{fig:number}.

                \begin{figure*}[]
                        \centering
                        \centering
                        \includegraphics[width=0.99\textwidth]{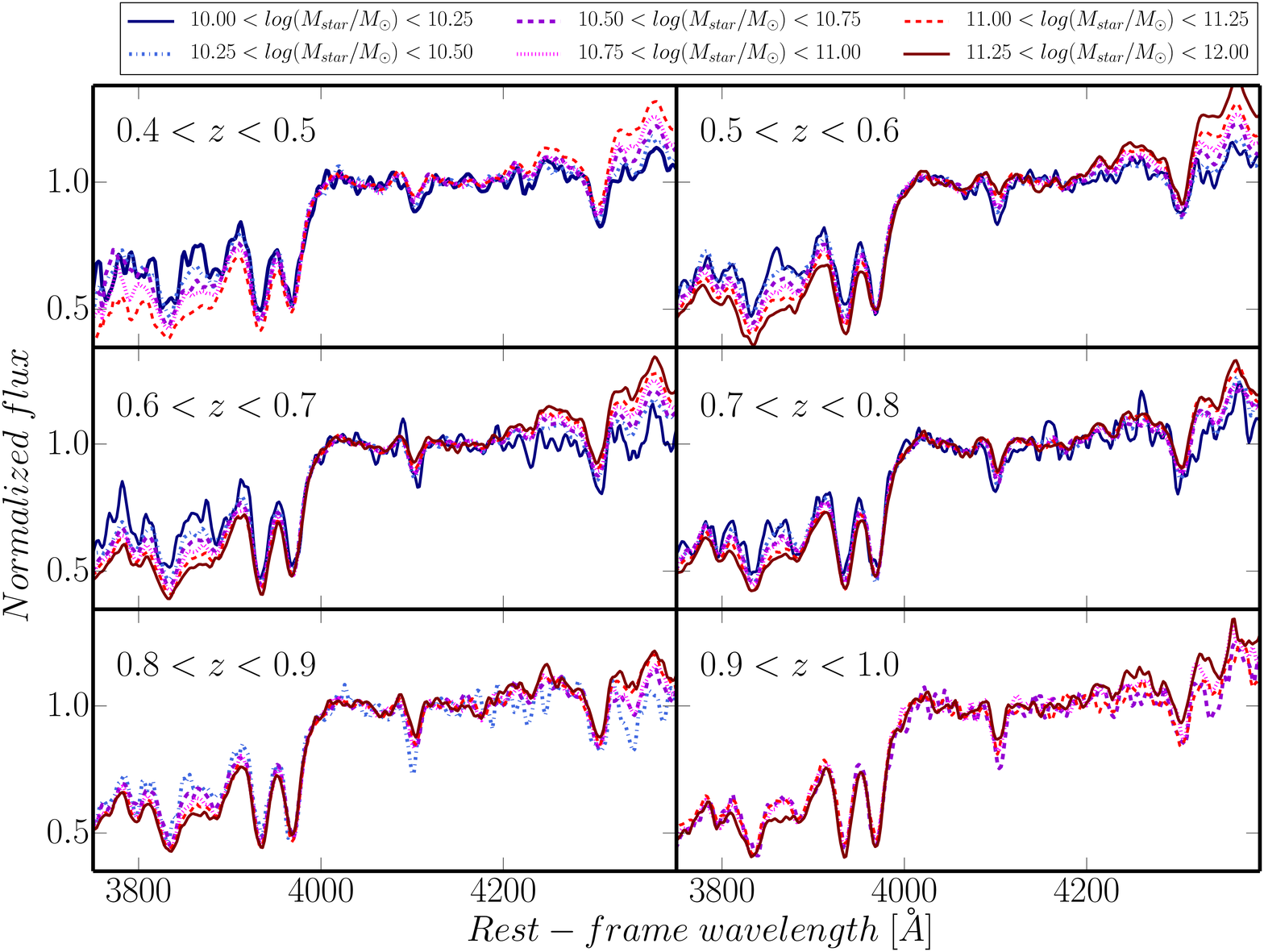}
                        \caption{Stacked spectra of VIPERS passive red galaxies in fixed redshift bins  and different stellar mass bins.  
                                Rest-frame spectra were normalized in the region $4010< \lambda < 4080$ and $4125< \lambda < 4200$ $\AA$.  }
                        \label{fig:stackedspectra}
                \end{figure*}
                
\begin{figure*}[]
                \centering
                \centering
                \includegraphics[width=0.99\textwidth]{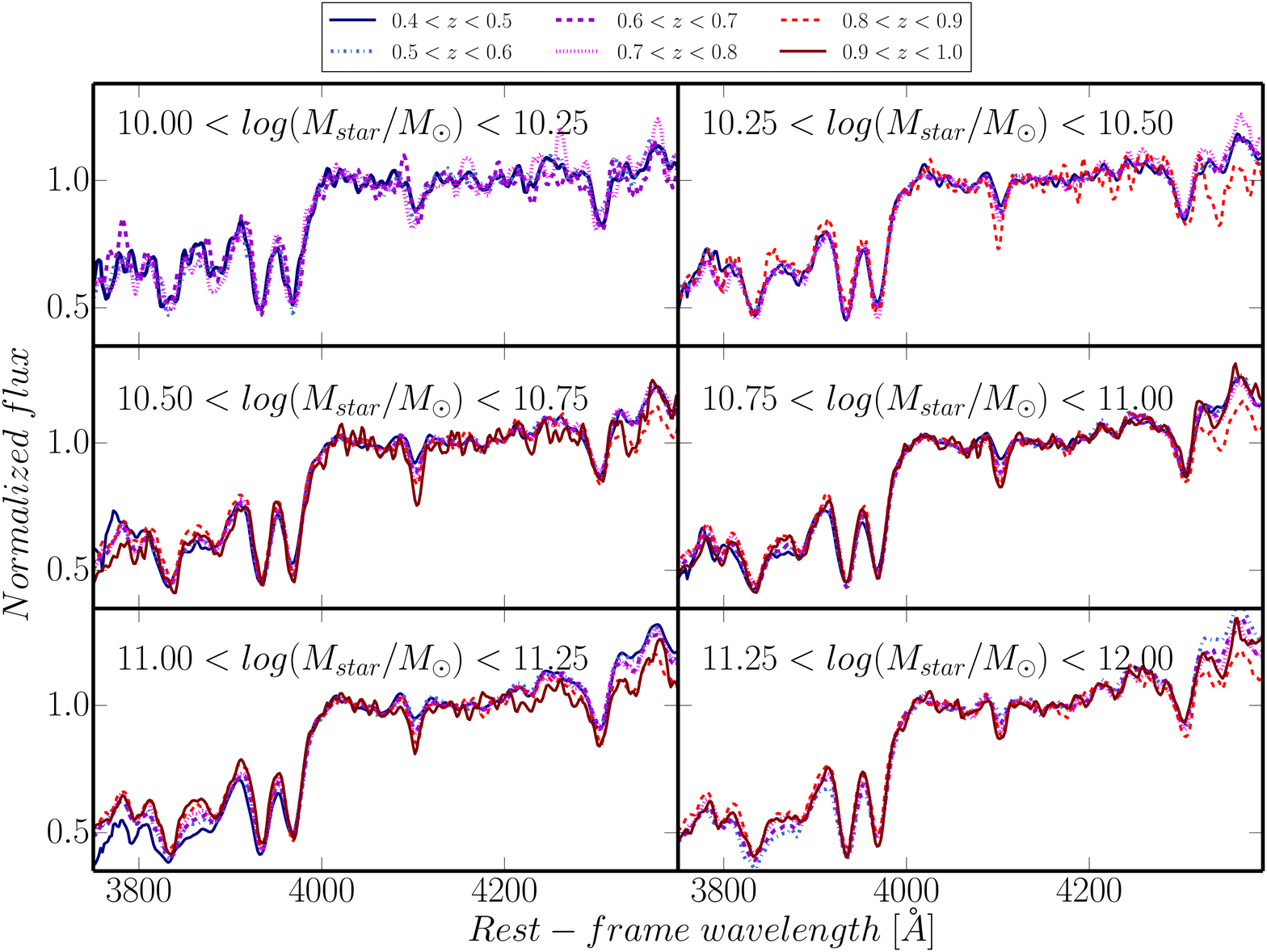}
                \caption{Stacked spectra of VIPERS passive red galaxies in fixed stellar mass bins  and different redshift bins.  
                        Rest-frame spectra were normalized in the region $4010< \lambda < 4080$ and $4125< \lambda < 4200$ $\AA$.  }
                \label{fig:stackedspectramass}
        \end{figure*}

        The VIPERS passive red galaxies stacked spectra are shown, limited to the wavelength range 3700 - 4600$\AA{}$, grouped by redshift bins, in Fig.~\ref{fig:stackedspectra}, and grouped by stellar mass bins in Fig.~\ref{fig:stackedspectramass}. Based on those two figures we can see 
        that there is considerably more variation among the spectra as a function of stellar mass (at fixed redshift, 
        Fig.~\ref{fig:stackedspectra}) than there is as a function of redshift (at fixed stellar mass, Fig.~\ref{fig:stackedspectramass}). 
        The low-mass passive red galaxies are much bluer and have stronger features around $\lambda \sim  3800\AA$ than the higher mass counterparts. 
        Although in this work we are focusing only on the $H\delta_{A}$ and $D4000_{n}$ features, we can see that the stacked spectra of VIPERS passive 
        galaxies contain significantly more information, which will be exploited in a future paper.

        \subsection{Spectral features measurement uncertainties}\label{sec:unc}
        
        Uncertainties on the spectral features measurements carried out on stacked spectra must include the effects of population variance
        inside each stellar mass and redshift bin, alongside the pure statistical uncertainty owing to the finite S/N ratio of the
        stacked spectra themselves. To include this population variance effect in our uncertainty estimates we used a Monte Carlo (MC) 
        approach. For each redshift and stellar mass bin we produced 500 different stacked spectra by drawing individual spectra from the 
        total dataset with the possibility of repetition, keeping the number of drawn spectra equal to the total number of individual spectra. 
        We then measured $D4000_{n}$ and $H\delta_{A}$ for each stacked spectrum, and obtained the standard deviation of the measurements for each spectral feature. Since this scatter due to population variance is always larger than the statistical uncertainty of the measurement
        on the stacked spectrum, we adopted the $1\sigma$ of the distribution of $H\delta_{A}$ and $D4000_{n}$ measurements on the MC stacks
        as the true value for the uncertainty of our measurements.

        \subsection{The SDSS comparison sample}\label{sec:sdss}
        
        To investigate evolutionary trends, we extend our redshift baseline including a local sample of passive red galaxies from the SDSS dataset. We used the DR12 CAS database\footnote{\url{http://skyserver.sdss.org/dr12/en/}}, from which we have extracted 
        photometric and emission line measurements, the values of stellar mass, and $H\delta_{A}$ and $D4000_{n}$ measurements for all galaxies 
        in the redshift range $0.15 < z < 0.25$. Spectral indicators were  measured on single rest-frame spectra using the same definitions as 
        for the VIPERS sample (see Sect. \ref{sec:sp}).  Stellar masses for the SDSS sample were estimated  using a Bayesian statistical approach 
        and based on a grid of models similar to that outlined by~\cite{kauffmana}. As spectra were measured through a 3 arcsec aperture, 
        the models were based only on the galaxy photometry, rather than on the spectral indices used originally  by~\cite{kauffmana}. A~\cite{kroupa} 
        IMF was adopted for the SDSS mass estimates, which may introduce a slight systematic mean offset with respect to the VIPERS stellar 
        mass scale for which a~\cite{chabrier} IMF was assumed\footnote{The scaling factor from a~\cite{chabrier} to a~\cite{kroupa} IMF is $\sim 1.1$~\citep{davidzon}.}.
        
        In order to apply the same photometric selection criteria to the SDSS sample as that used for the VIPERS passive galaxy sample, we first \textit{k} corrected the $ugriz$ SDSS model magnitudes. Magnitudes are estimated using an independent galaxy model (the better of the exponential or de Vaucouleurs fits in the r band) and convolved with the band-specific point spread function (PSF). The model magnitudes are recommended to use for measuring galaxies colors, which should not be biased in the absence of color gradients. We used \textit{k }corrections provided by the SDSS Photo-z catalog in DR12, which are based on template fitting methods as described in~\cite{oyaizu}. Then we applied a correction for
        Galactic extinction computed following~\cite{schlegel} and finally we derived absolute $U$ and $V$ magnitudes in the Vega system for each galaxy, following the conversion recipes provided 
        by~\cite{jester}. By applying the \cite{fritz} $U-V$ color selection criterion discussed in Sect.~\ref{sec:etg} to the sample
        of SDSS galaxies described above, we obtained a sample of 72,810 SDSS passive red galaxies. A further check on the presence of
        significant $\rm{H\alpha}$ and $\rm{[OII]\lambda3727}$ equivalent width in the galaxy spectrum ($\rm{EW(H\alpha)} < 5\AA$ and $\rm{EW([OII]\lambda3727} < 5\AA$) 
        has lead to the rejection of 62 galaxies, for a final SDSS comparison sample composed of 72,748 galaxies, 
        which we can use as local counterpart of the VIPERS sample. 

        We did not build stacked SDSS spectra because the S/N ratio of individual SDSS spectra is large enough to provide reliable $D4000_{n}$ and $H\delta_{A}$ measurements for the single galaxies. Therefore we simply obtained a linear fit for the $D4000_{n}$ 
        and $H\delta_{A}$ versus stellar mass relations to be used as a comparison with similar relations obtained for the VIPERS
        passive red galaxies.

        \section{Results}\label{sec:results}
        
        \subsection{Dependence of spectral features on stellar mass}\label{massrelation}
        
\begin{figure*}[ht]
                
                \begin{subfigure}[a]{0.49\textwidth}
                        \includegraphics[width=\textwidth]{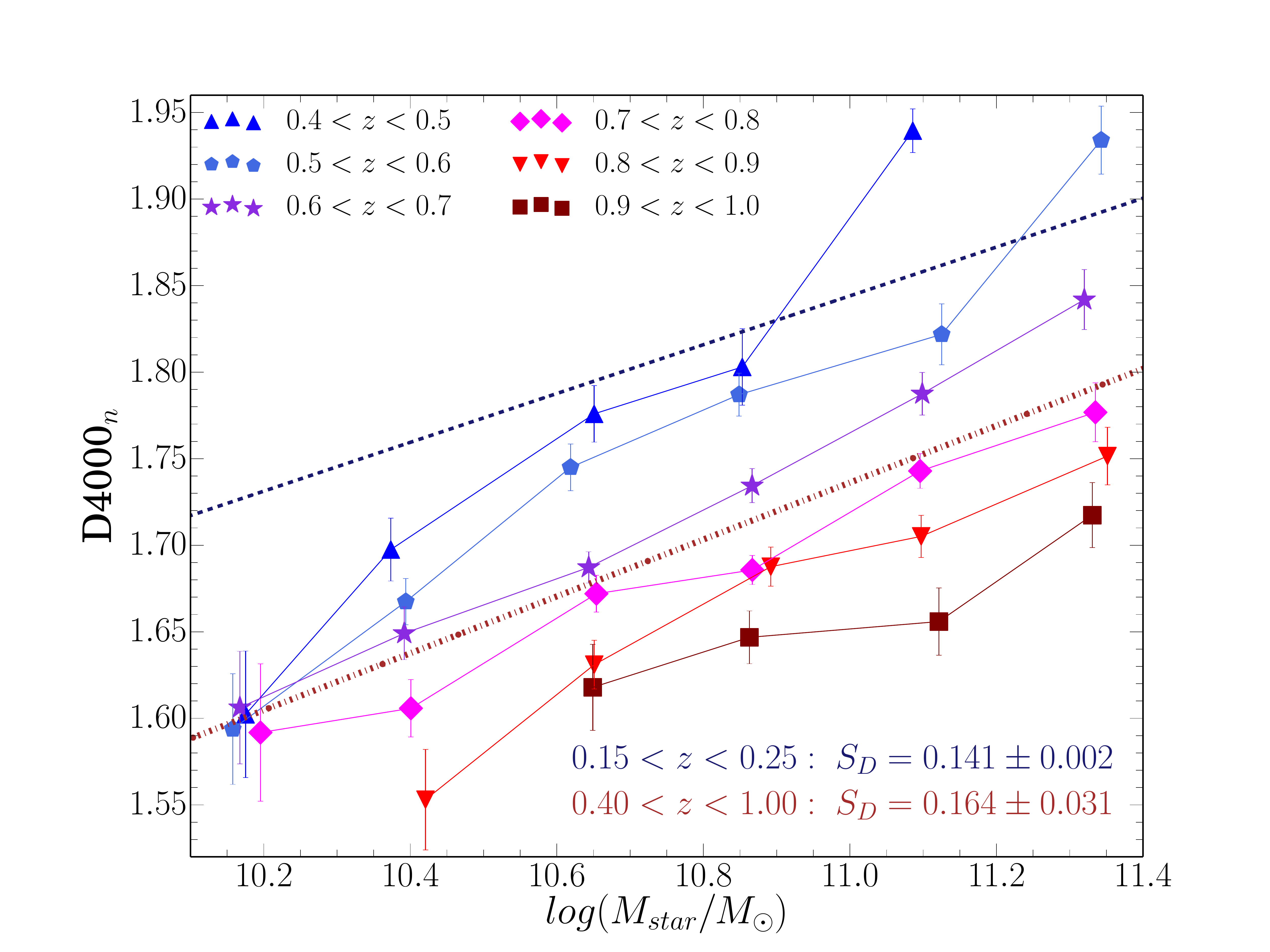}
                        \caption{$D4000_{n}$-stellar mass relation;}
                        \label{fig:d4000mass}
                \end{subfigure}
                \begin{subfigure}[a]{0.49\textwidth}
                        \includegraphics[width=\textwidth]{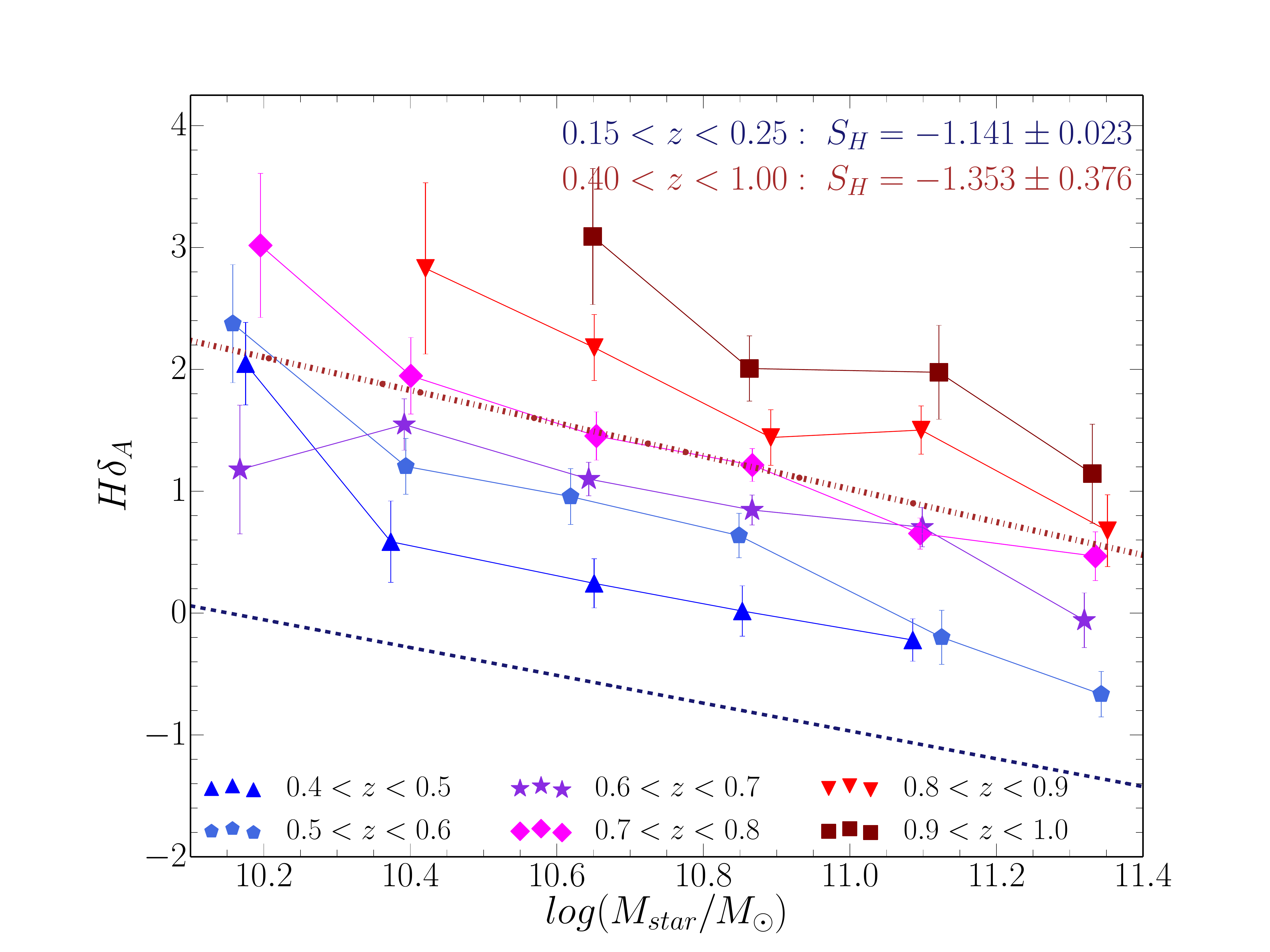}
                        \caption{$H\delta_{A}$-stellar mass relation.}
                        \label{fig:hdelta}
                \end{subfigure}
                \caption{$D4000_{n}$ and $H\delta_{A}$ as a function of stellar mass for VIPERS stacked spectra of passive red galaxies. 
                        The linear fits to the whole VIPERS sample and to the SDSS sample are shown as brown dot-dashed and blue dashed lines, respectively. The slopes and relative uncertainties of the linear relationships between spectral features $D4000_{n}$ and $H\delta_{A}$ ($S_{D}$ and $S_{H}$), respectively, and the stellar mass for the local SDSS sample and for the VIPERS sample are annotated.  }
                \label{fig:lick}
        \end{figure*}

Before examining the age difference between low- and high-mass passive red galaxies, we look first at the change of spectral features  as a function of stellar mass. 
        The $H\delta_{A}$ and $D4000_{n}$ measurements derived from the VIPERS stacked spectra are plotted as a function of stellar mass 
        in Fig.~\ref{fig:lick}, where the stellar mass value being plotted is the median value for all galaxies within each bin described 
        in Sect.~\ref{sec:stacking}. The error bars are taken from Monte Carlo simulations described in Sect.~\ref{sec:unc}.
        In the same figure we also plot the linear fit for $H\delta_{A}$ and $D4000_{n}$ as a function of stellar mass for the SDSS 
        passive red galaxies comparison sample (blue dashed line),

        A strong dependence of the two spectral indicators on both stellar mass and redshift is clearly seen in the two plots.
        Passive galaxies at lower redshift have a stronger $D4000_{n}$  and a weaker $H\delta_{A}$ with respect to galaxies at higher 
        redshift within any stellar mass bin. This evolution in spectral features implies that, as expected for a population of
        passively evolving galaxies, the lower redshift stellar populations are older than their higher redshift counterparts. A detailed analysis of the epoch when stars formed in these systems is presented in Sect.~\ref{sec:downsizing}.
        At the same time massive passive red galaxies have a stronger $D4000_{n}$ break and a weaker $H\delta_{A}$ than the lower mass galaxies, 
        within any redshift bin, with both spectral indicators varying almost linearly as a function of stellar mass.
        
        The highest stellar mass bins in the redshift range $0.4 < z < 0.6$ tend to have high $D4000_{n}$ values
        well above the value one would expect based on the observed trends at lower masses or higher redshift (see Fig.~\ref{fig:d4000mass}). 
        We do not have any obvious reason to reject these data points, other than to notice that the number of galaxies within these
        two specific bins is very small. Therefore any currently undetected incompleteness that might be present in our sample would
        affect these small samples comparatively more than all the other, more richly populated, bins (see Sect.~\ref{sec:etg}).

        The slopes of $D4000_{n}$ and $H\delta_{A}$- stellar mass relations ($S_{D}$, $S_{H}$, respectively) do not change significantly in the VIPERS redshift range. Thus, we combine the full VIPERS sample to derive single slope values for the $D4000_{n}$ and $H\delta_{A}$- stellar mass relations.

        The $D4000_{n}$-stellar mass relation found for VIPERS galaxies is in agreement with that found for zCOSMOS ETGs at a 
        similar redshift~\citep{morescoage}. The mean difference in $D4000_{n}$ between stellar mass 10.2 < $\rm{log(M_{star}/M_{\odot})}$ < 10.8 and redshift range $0.4 < z < 1.0$ 
        (to be consistent with the~\cite{morescoage} computation) is in fact almost identical to that found for zCOSMOS 
        (0.11$\pm$ 0.02, 0.10 $\pm$ 0.02, respectively). 
        
        Taking advantage of the size of the VIPERS sample, we can extend our internal comparison to higher stellar masses and lower redshift.  
        The mean difference in $D4000_{n}$ for the whole stellar mass (between $\rm{log(M_{star}/M_{\odot})\sim 10.18}$, and $\rm{log(M_{star}/M_{\odot})\sim 11.34}$) and redshift ($0.4 < z < 1.0$) ranges is equal to $0.19 \pm 0.04$ and is very well matched with the mean difference 
        we can deduce from the linear fit computed for the SDSS sample ($0.17 \pm 0.01$).
        
        The dependence of spectral features on mass does not show any significant evolution from $z\sim 1$ to $z\sim 0$. Both slopes are consistent within error bars with those we obtained for the SDSS comparison sample 
        ($S_{D} =0.141 \pm 0.002$, $S_{H}=-1.141 \pm 0.023$, blue dashed lines in Fig.~\ref{fig:lick}). A quantitative comparison between slopes of $D4000_{n}$ and $H\delta_{A}$- stellar mass relations for low- and high-redshift red passive galaxies is presented in Appendix~\ref{app:B}.

        Our results confirm the dependence of the 4000$\AA$ break on stellar mass among red passive galaxies first discussed for galaxies in the local Universe
        by~\cite{kauffmana} and later observed for galaxies in the redshift range $0.45 < z < 1.0$ by \cite{morescoage}. Now, taking advantage of the large VIPERS dataset,  we extended these results. Here we can see       with good accuracy how this dependence remains basically unchanged over the whole redshift range $0 < z <1$, except 
        for the global shift (as a function of cosmic time) toward higher $D4000_{n}$ values introduced by the aging of the passively evolving 
        red passive galaxy stellar populations. A new result of this work is that this evolution is completely mirrored by the evolution of 
        the $H\delta_{A}$ index, which becomes weaker with increasing stellar mass and also shows the signature of a passive 
        evolutionary scenario over the whole redshift range $0 < z <1$.
        
        As the main driver for the evolution of both spectral indices is the aging of stellar populations in galaxies~\citep[see][]{kauffmana},        we can conclude that we find clear indications that, on average, passive red galaxies with lower masses have younger mean ages. Before we consider this 
        as a strong confirmation of the downsizing scenario, in which high-mass passive red galaxies are populated by older stellar populations 
        than low-mass galaxies, we also need to consider the possible effect of changes in metallicity on the $D4000_{n}$ evolution, which we
        discuss in Sec.~\ref{sec:downsizing}.

        \subsection{The $H\delta_{A}$ - $D4000_{n}$ relation}
        
\begin{figure*}[ht!]
                \centering
                \centering
                \includegraphics[width=0.9\textwidth]{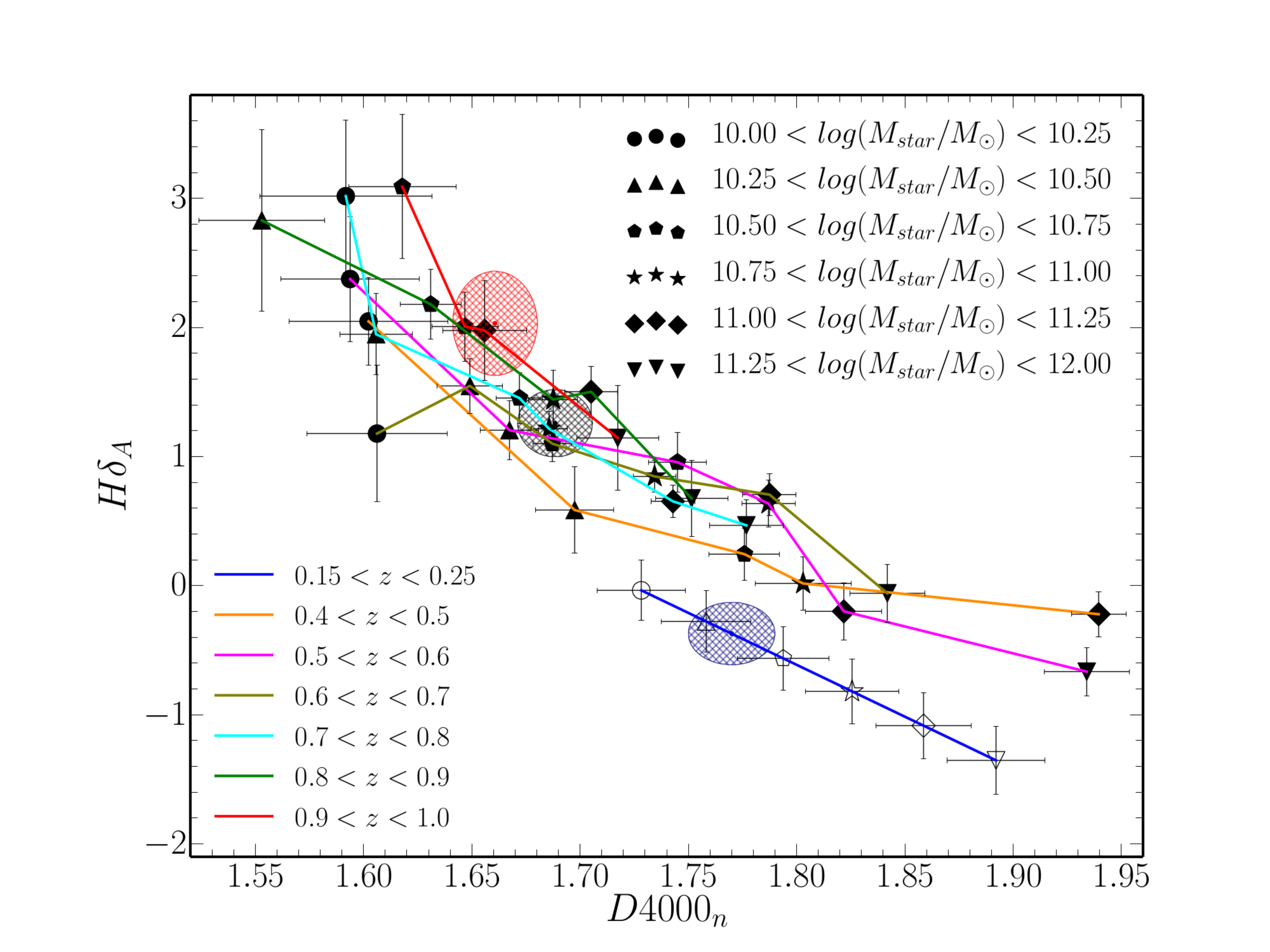}
                \caption{$H\delta_{A}$ as a function of  $D4000_{n}$ for different redshift and stellar mass bins obtained for VIPERS stacked spectra. 
                        Error bars were obtained on the basis of Monte Carlo simulations.  
                        Values of $H\delta_A$ for SDSS passive red galaxies obtained from  the linear fit of $H\delta_A$-stellar mass relation for median stellar mass bins.  $D4000_{n}$ values were obtained in an analogous way. Error bars were derived from the uncertainties of the linear fits. 
                        The spectral feature strength expected for galaxies with stellar mass corresponding to the transition mass at which the mass functions of the star-forming and quiescent galaxy populations cross is indicated with blue, black, and red ellipses at $z \sim 0.1$, $z \sim 0.7$, $z \sim 1$, respectively. The error bars of the estimated spectral
                        features for transition masses correspond to the areas of ellipses.
                }
                \label{fig:lick2}
        \end{figure*}

        In the previous section we discussed how the two spectral indices, $D4000_{n}$ and $H\delta_{A}$, qualitatively delineate the same
        evolutionary scenario for the VIPERS passive red galaxies population, both as a function of stellar mass and redshift.
        Here we directly examine the connection between the two indices, which appear to be very tightly correlated over the full range of stellar 
        masses and redshifts covered by the VIPERS sample, as shown in Fig.~\ref{fig:lick2}. A small offset is instead present between the VIPERS 
        and the SDSS data, which is probably due to a combination of the different methods of measuring the indices (VIPERS stacked spectra with respect to to SDSS single spectra) and of the further aging of the stellar population between the VIPERS lowermost redshift bin
        and the SDSS mean redshift. 
        
        This correlation is consistent with previous low-redshift results~\cite[e.g.,][]{kauffmana, borgne}: low-mass passive red galaxies have 
        weaker $D4000_{n}$ and stronger $H\delta_{A}$ than their high-mass counterparts. This implies that they are dominated by relatively 
        younger stellar populations, which are the result of a more recent star formation activity; this is discussed further in 
        Sect.~\ref{sec:downsizing}. We can also observe a significantly larger scatter in $H\delta_{A}$ at fixed $D4000_{n}$ when the 
        $D4000_{n}$ value is below approximately 1.65 to 1.7. The value of stellar mass that is associated with this range of $D4000_{n}$ 
        values, within the redshift bins we are sampling with our data (see Figs.~\ref{fig:lick} and \ref{fig:lick2}), corresponds very closely to so-called 
        transition mass, where the mass functions of the star-forming and quiescent galaxy populations cross (above the transition mass 
        quiescent galaxies are the most numerous population, while below that mass star-forming galaxies become dominant). In the local
        Universe the transition mass has been estimated to be $3\cdot 10^{10}$ $\rm{M_{\odot}}$~\citep[][shown with a blue dashed ellipse in Fig.~\ref{fig:lick2}]{kauffmana, bell2007}, but this value has been shown to increase with redshift~\citep{bundy,pannella,muzzin,davidzon}, reaching $\sim 10^{11}$ $\rm{M_{\odot}}$ at $z \sim 1$~\citep[][indicated with a red dashed ellipse in Fig.~\ref{fig:lick2}]{pannella}. The axes of these ellipses correspond 
        to the error bars of the estimates for the two spectral features in the stellar mass bin coinciding with the transition mass at 
        the given redshift. 
        
        We consider this connection between the scatter in the $H\delta_{A}$ versus $D4000_{n}$ relation and the transition mass as evidence
        of a specific evolutionary pattern for the passive red galaxies population. Above the transition mass, where no significant contamination 
        from recently quenched massive star-forming galaxies is to be expected for the passive red galaxies population, they seem to form a rather 
        homogeneous population with their evolution restricted to a pure passive evolution of their stellar populations. Below the transition
        mass, where instead some contamination from recently quenched star-forming galaxies is to be expected, the passive red galaxies population
        appears to be less homogeneous with the $H\delta_{A}$ index indicating the presence of a larger range of star formation histories.
        
        We find further evidence for this scenario by analyzing the scatter in $H\delta_{A}$ index values within single stellar mass and 
        redshift bins. We divided the VIPERS passive galaxy samples belonging to the lowest and highest mass bin into five subsamples, 
        sorted on the basis of the individual spectrum $D4000_{n}$ measurements, and then we obtained a stacked spectrum for each subsample, measured 
        the $H\delta_{A}$ index on these stacked spectra, and compared the scatter in the measurements between the lowest and highest mass sets.
        The result is that within the lowest mass passive red galaxies the scatter in $H\delta_{A}$ index is 30 to  40\% larger than what is 
        observed within the highest mass counterparts, which is a clear indication that the low-mass passive red galaxies form a population that is
        significantly less homogeneous than that composed by the high-mass galaxies.

        \subsection{Metallicity dependences}\label{sec:metallicity}
        
        Before we can confidently claim that our results, discussed in the previous two sections, are to be considered a\ confirmation of 
        the downsizing scenario, providing evidence that massive galaxies have older stellar populations than less massive galaxies all the way
        up to $z \sim 1$, we need to consider the possible effects on those results of a systematic change in the metallicity of the stellar populations as 
        a function of galaxy stellar mass.
        
\begin{figure*}[ht!]
                
                \begin{subfigure}[a]{0.49\textwidth}
                        \includegraphics[width=\textwidth]{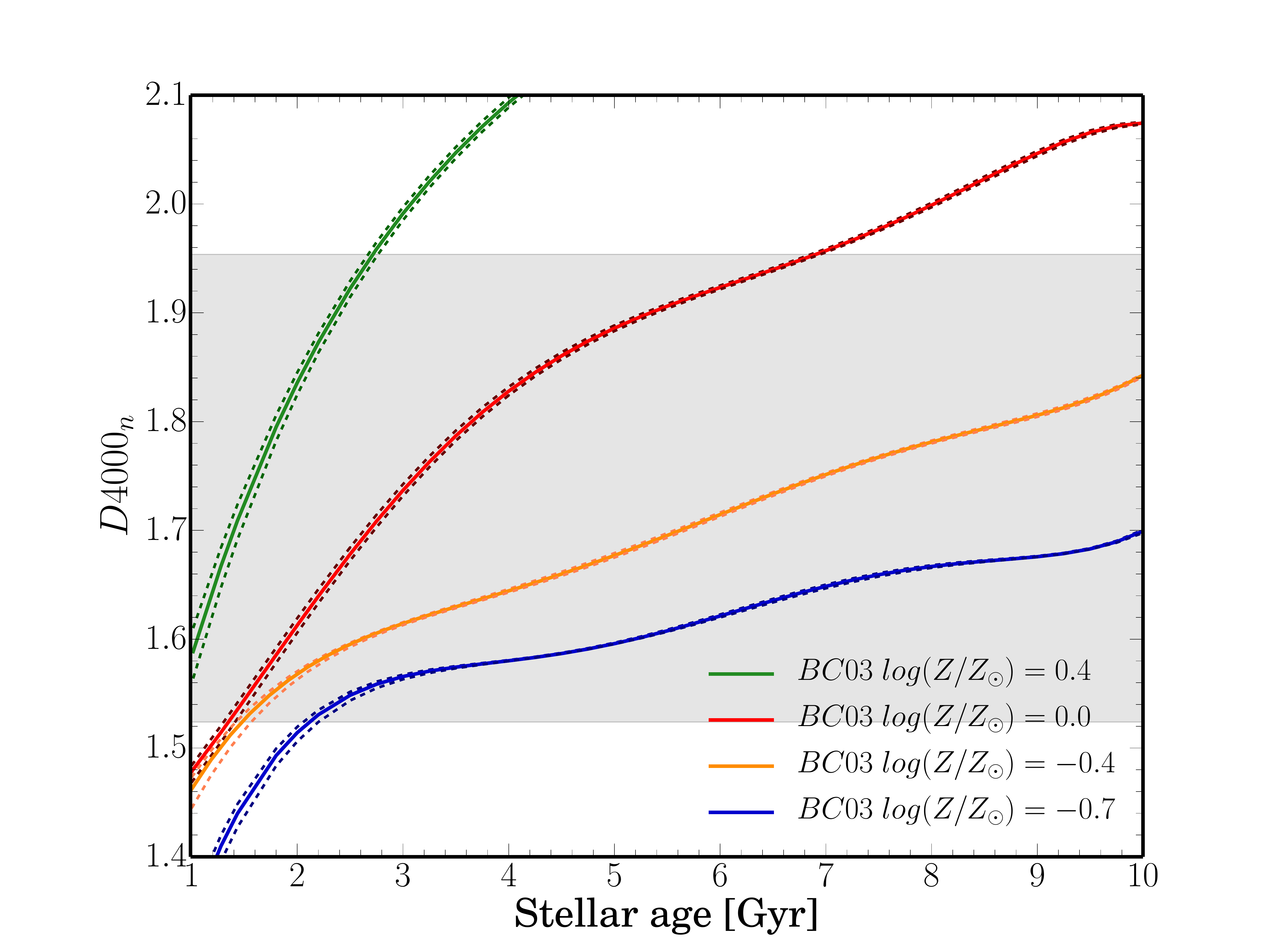}
                        \caption{$D4000_{n}$-stellar age relation;}
                        \label{fig:d4000}
                \end{subfigure}
                \begin{subfigure}[a]{0.49\textwidth}
                        \includegraphics[width=\textwidth]{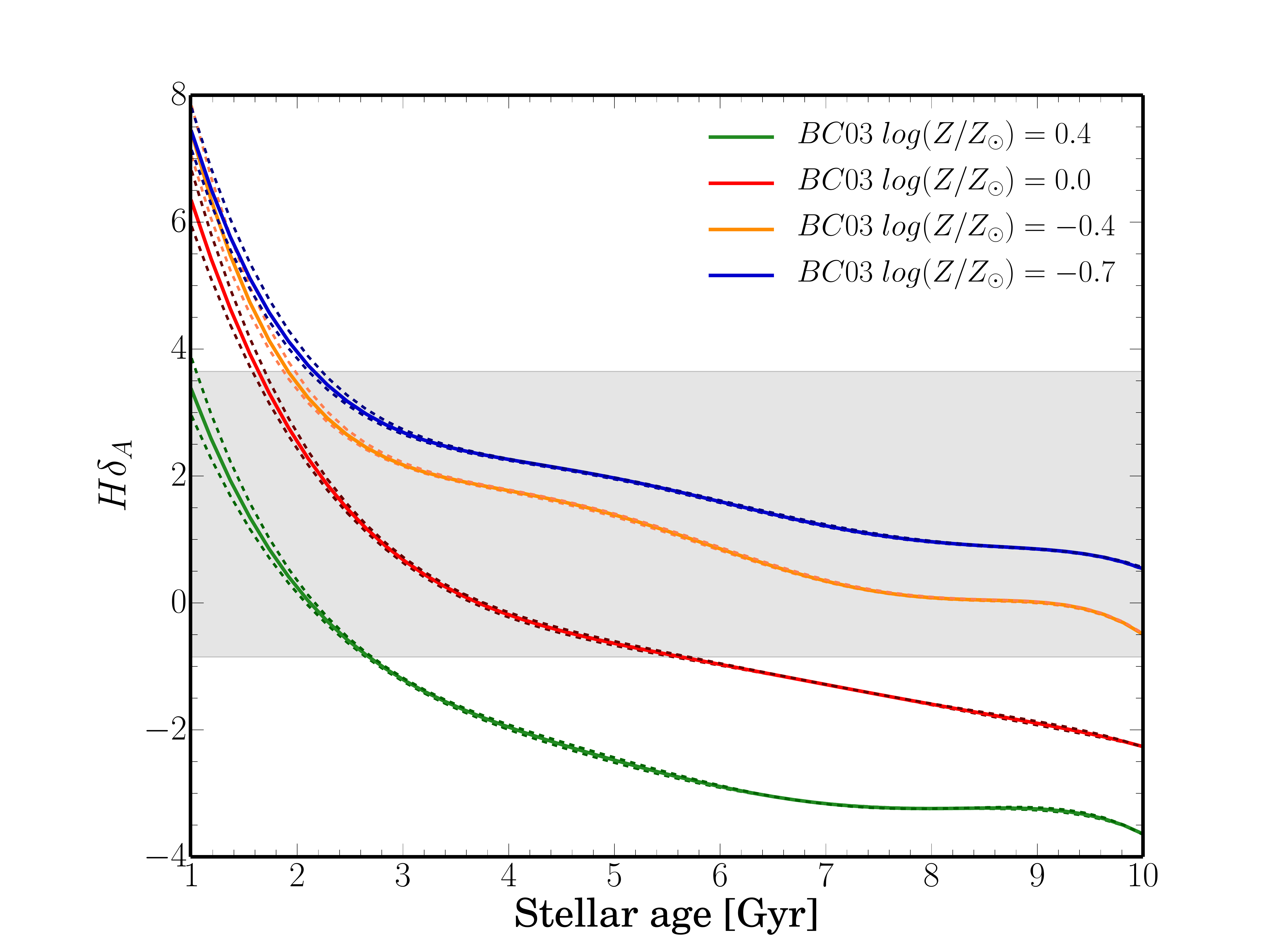}
                        \caption{$H\delta_{A}$-stellar age relation.}
                        \label{fig:hdelta}
                \end{subfigure}
                \caption{$D4000_{n}$ and $H\delta_{A}$ as a function of stellar age derived from a grid of synthetic spectra (BC03 model) are plotted assuming stellar metallicity: $\rm{log(Z/Z_{\odot})=0.4}$, $\rm{log(Z/Z_{\odot})=0.0}$, $\rm{log(Z/Z_{\odot})=-0.4}$, $\rm{log(Z/Z_{\odot})=-0.7}$. The model assumes Chabrier IMF and one single stellar burst with timescales $\tau = 0.1, 0.2, 0.3$ Gyrs. Dashed lines correspond to $\tau =$ 0.1, and 0.3 Gyrs, while the solid ones to $\tau =$ 0.2 Gyr. Gray areas correspond to the ranges of $D4000_{n}$ and $H\delta_{A}$ obtained for VIPERS stacked spectra of passive red galaxies.  }
                \label{fig:age}
        \end{figure*}

        To this purpose, we estimated the influence of the variation in stellar metallicity from the comparison of 
        our spectral features measurements with those derived for a grid of synthetic galaxy spectra, based on BC03 models covering a range of metallicities. The synthetic spectra were generated using the Padova 1994 stellar evolutionary tracks 
        and the high-resolution STELIB spectral library with a star formation history assumed to be a single burst with a timescale 
        $\tau = 0.1, 0.2, 0.3$ Gyr. For each value of $\tau$, a set of synthetic spectra was obtained for stellar  ages in the range from 1 to 
        10 Gyr, with steps of 0.25 Gyr. For the grid of model spectra, we calculated  $D4000_{n}$ and $H\delta_{A}$ using the same 
        definitions and tools discussed in Sect. \ref{sec:sp}. We then obtained the nominal $D4000_{n}$ and $H\delta_{A}$-stellar age relations 
        based on these measurements for the BC03 models with metallicities, $\rm{log(Z/Z_{\odot})=0.4}$, $\rm{log(Z/Z_{\odot})=0.0}$, $\rm{log(Z/Z_{\odot})=-0.4}$, $\rm{log(Z/Z_{\odot})=-0.7}$~(see Fig.~\ref{fig:age}), 
        
\begin{figure}[ht!]
                        \centering
                        \centering
                        \includegraphics[width=0.44\textwidth]{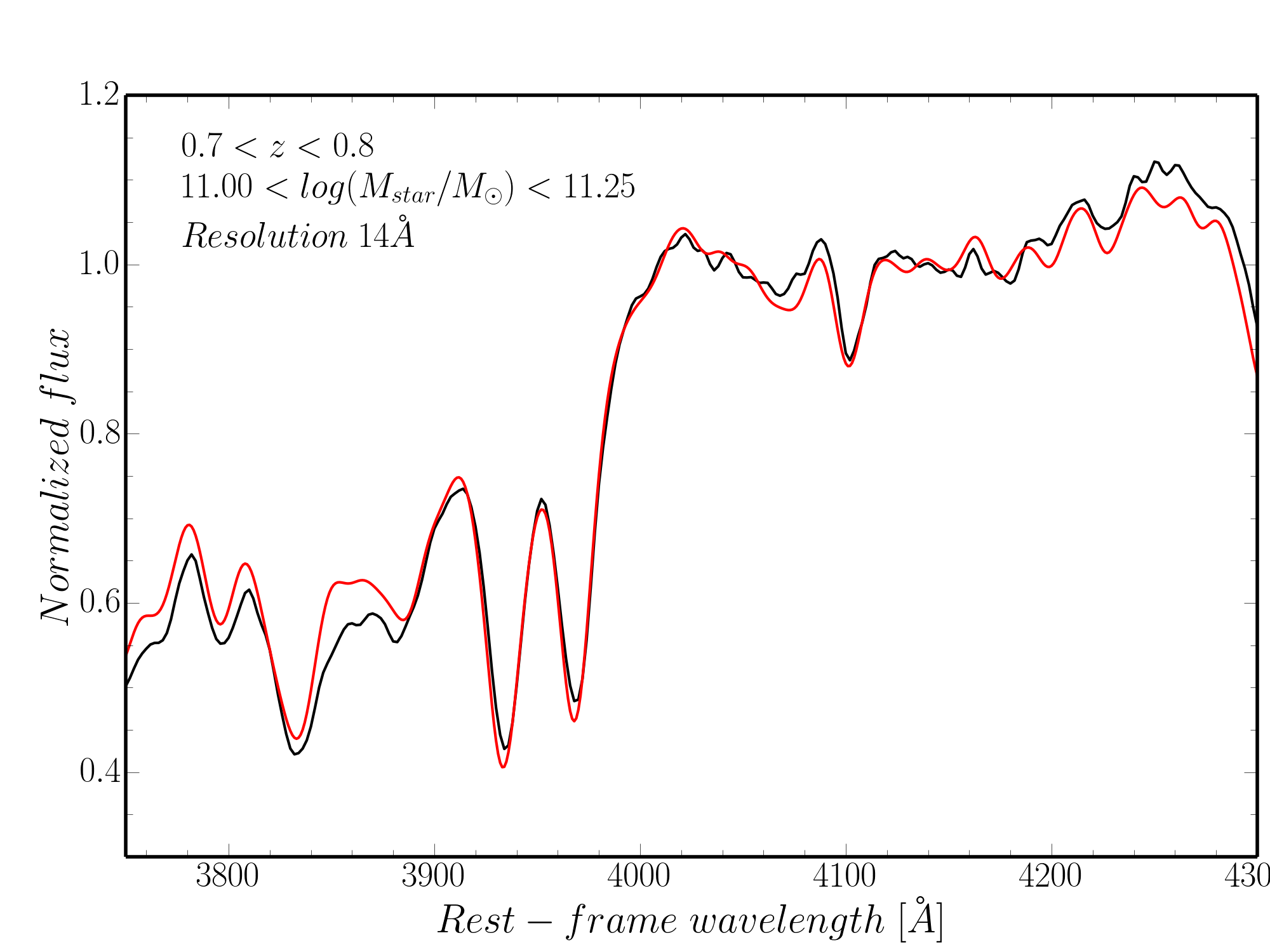}
                        
                        \caption{VIPERS exemplary stacked spectrum is shown as black line. A synthetic spectrum with a single burst of duration $\tau=0.2$ Gyr, solar metallicity, and age of 3 Gyr is overplotted in red. The high-resolution model was downgraded to the resolution of $\sim 14\AA$.   }
                        \label{fig:model}
                \end{figure}

                Spectral features (especially $H\delta_{A}$) are sensitive to the spectral resolution of the spectra. Thus, we matched the resolution between the model and the real VIPERS spectra. The resolution of the BC03 models
                we used is $3\AA$ across the wavelength range from $3200\AA$ to $9500\AA$. This is significantly higher than that of the VIPERS spectra that were obtained using the
                VIMOS LR-Red grism, which provides a resolving power of approximately 230. For our target galaxies, covering the redshift range from 0.5 to 1 with spectral 
                features observed in the wavelength range from 6000 to 8500\AA; this translates into a resolution for the rest-frame spectra ranging from 13 to 16\AA.
                We have chosen therefore to downgrade the resolution of the BC03 models to a common value of 14\AA, which provides an excellent match to our stacked spectra,  
                as shown in Fig.~\ref{fig:model}, where we have overplotted a synthetic spectrum with the age of 3~Gyr on a VIPERS stacked spectrum .

        To estimate the variations of stellar metallicity as a function of galaxy stellar mass we used the work of~\cite{gallazzi2014}, who have estimated a mean metallicity for passive high-mass ($10^{11}$ $\rm{M_{\odot}}$) galaxies  
        of $\rm{log(Z/Z_{\odot})=0.07\pm 0.03}$, at a mean redshift $z=0.7$ with a relatively flat slope of the metallicity versus stellar mass relation ($0.11\pm0.10$).
        
        Considering fixed metallicity at the level of solar metallicity, the observed change in $D4000_{n}$ as a function of stellar mass, shown in Fig~\ref{fig:d4000mass},  would correspond to the mean age difference 
        between the high- and low-mass VIPERS passive red galaxies of approximately 2~Gyr. On the other hand, the mean change of metallicity for the highest stellar mass bin on the level of  0.1 in $\rm{log(Z/Z_{\odot})}$ would result in the expected change of $D4000_{n}$ equal to $\pm 0.06$ for galaxies with stellar age $\sim 4$~Gyr. According to~\cite{gallazzi2014}, the slope of the stellar metallicity-mass relation for passive galaxies equals to $0.15\pm0.03$ and $0.11\pm0.10$ at z=0.1, and z=0.7, respectively. Such a slope, and change in $D4000_{n}$ , give a predicted slope of $D4000_{n}$-mass relation on the level of 0.07, and 0.10, again for z=0.1, and z=0.7, respectively. These values compose a significant fraction of the slope measured for the VIPERS sample ($S_{D}$ = $0.164 \pm 0.031$, see Sect.~\ref{massrelation}). Thus, we can conclude that the $D4000_{n}$-mass relation is changing because of the variation both in the age and metallicity of stellar populations in passive red galaxies, however, we are not able to clearly distinguish both effects. 
        We do not consider significant for our results the small evolution of stellar metallicity with redshift that we can
        expect within the redshift range covered by our sample, since within these passive red galaxies the bulk of the stars were already formed at higher redshifts. This implies 
        no significant changes of metallicity in comparison to the local Universe~\citep[e.g.,][]{carson, gallazzi2014}. Moreover, the solar metallicity was also observed for quiescent galaxies at $z = 2$~\citep{toft}. Thus, we assume the stellar metallicity evolving with mass at all redshift bins. We expect solar metallicity for our sample up to $10^{11}$ $\rm{M_{\odot}}$ and a change on the order of 0.07 and 0.1 $\rm{log(Z/Z_{0})}$ for the higher mass bins of $\sim10^{11.1}$, $\sim10^{11.4}$ $\rm{M_{\odot}}$, respectively.

        \subsection{Signatures of downsizing}\label{sec:downsizing}
        
        We estimated the epoch of the last starburst (redshift of formation, hereafter: $z_{form}$) for the VIPERS passive red galaxies, based on the spectral feature measurements carried out
        on the VIPERS stacked spectra. The measured values were compared with the values obtained from the BC03 models with a Chabrier IMF and SFH assuming a single burst (with the same assumptions as in Sect.~\ref{sec:metallicity}, i.e., with $\tau$ =0.1,0.2, 03 Gyr and  metallicity evolving with mass). 
        As $z_{form}$ is calculated from the $D4000_{n}$ and $H\delta_{A}$-age relations obtained from BC03 models with a single burst of star formation, its value
        corresponds to the representative approximate epoch of the last major star formation episode within our passive red galaxies population. As such, it does not carry any information on the actual formation epoch. The real redshift of formation and SFH for
        these galaxies are likely to be more complex than a single burst at $z_{form}$, however, we adopt this definition of galaxy formation following the standard practice
        used in most recent studies on this subject~\citep[e.g.,][]{onodera2015, jorg2013}.

\begin{figure*}[ht!]
                \begin{subfigure}[a]{0.49\textwidth}
                        \includegraphics[width=\textwidth]{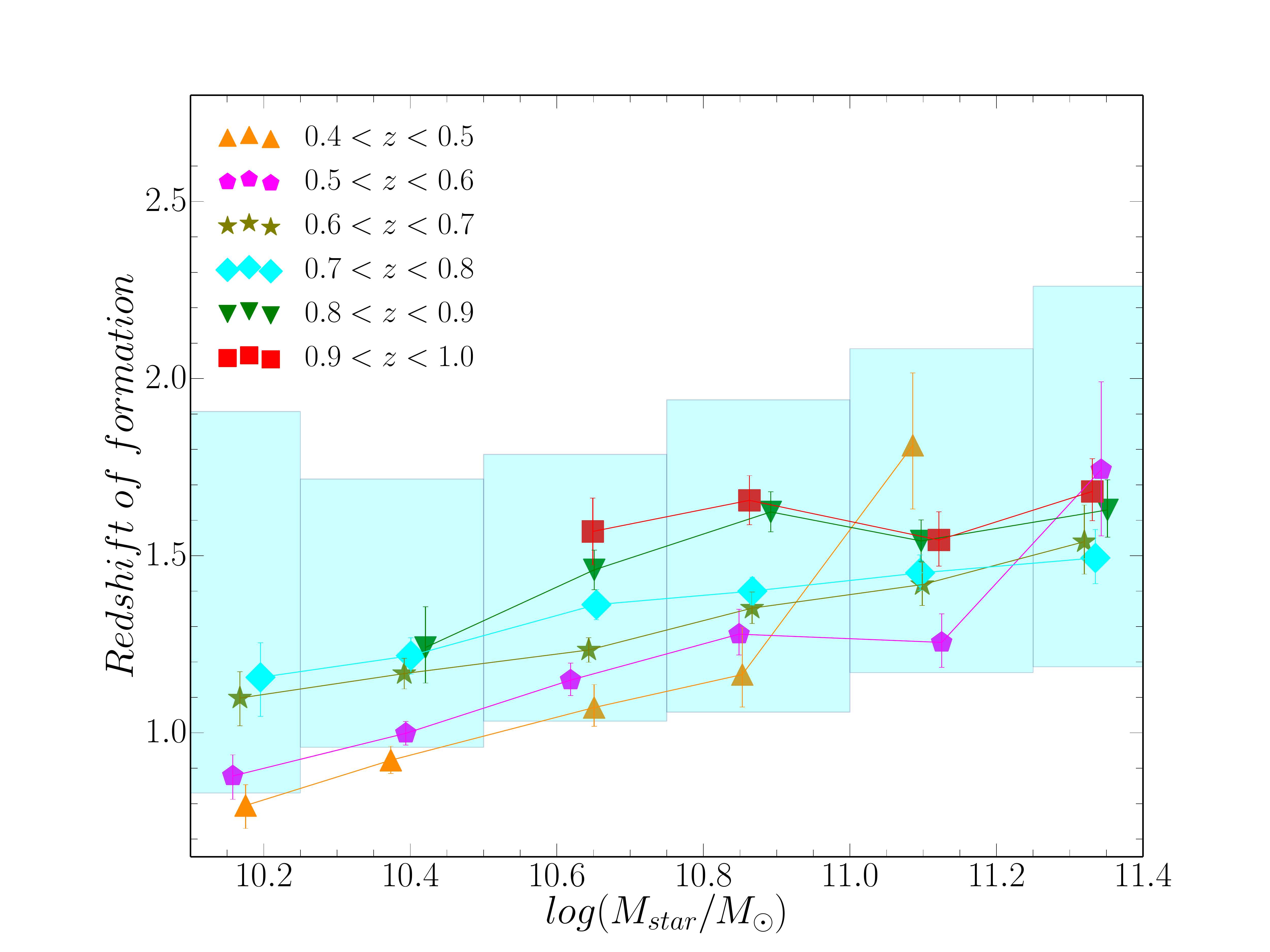}
                        \caption{based on $D4000_{n}$;}
                        \label{fig:rofd4000}
                \end{subfigure}
                \begin{subfigure}[a]{0.49\textwidth}
                        \includegraphics[width=\textwidth]{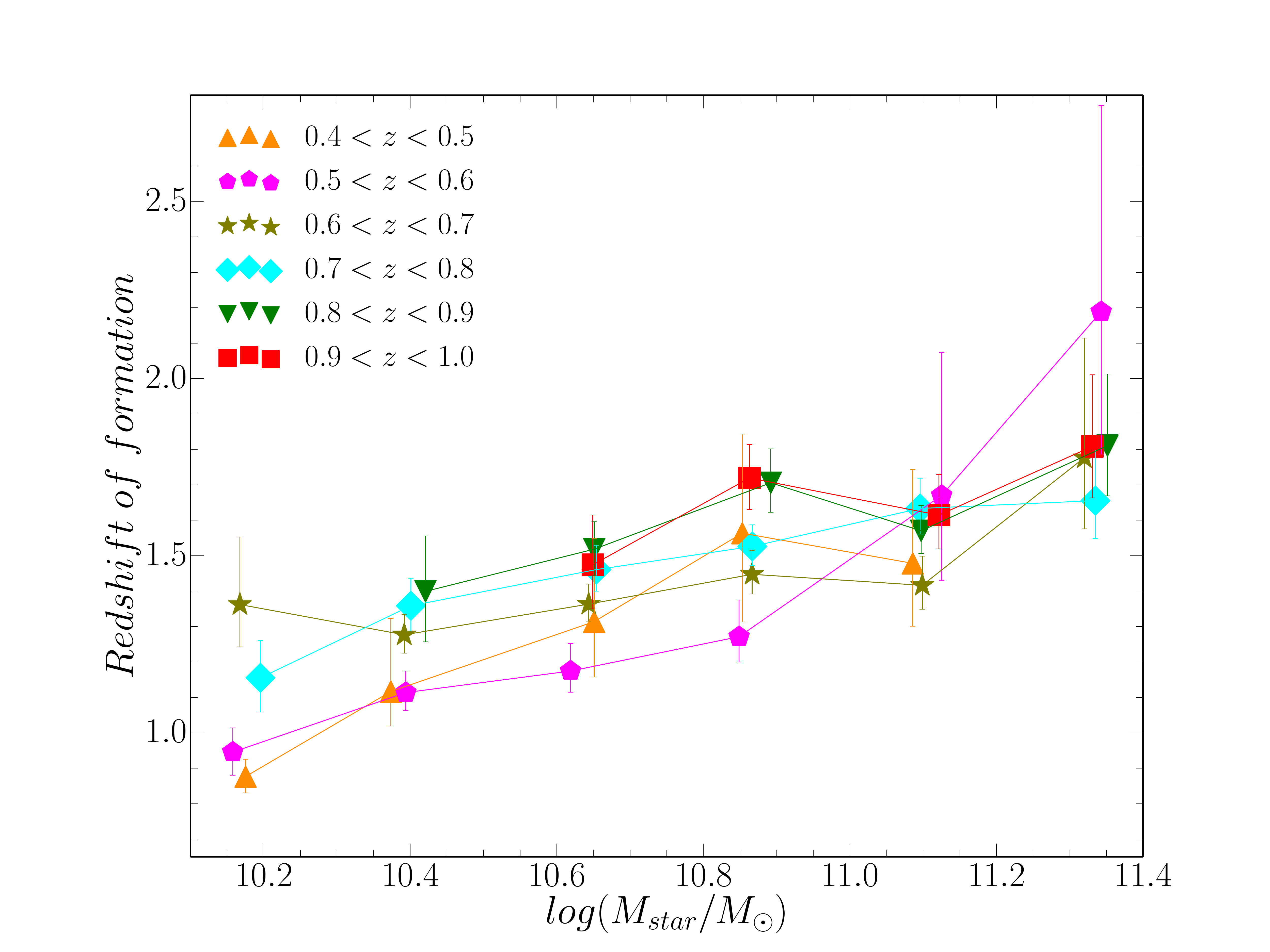}
                        \caption{based on $H\delta_{A}$.}
                        \label{fig:rofhdelta}
                \end{subfigure}

                \caption{Redshift of formation as a function of stellar mass for a sample of VIPERS passive red galaxies. The error bars correspond to different burst duration ($\tau = 0.1$ and 0.3 Gyrs).  
                        Both calculations were based on the BC03 model assuming stellar metallicity evolving with mass (0.1 $log(Z/Z_{\odot})$ for the highest stellar mass bin)}. The cyan areas correspond to the $1\sigma$ weighted distribution of redshift of formation obtained for individual spectra for redshift bin $0.7 < z < 0.8$ and different stellar mass bins. 
                \label{fig:rof}
        \end{figure*}

        In Fig. \ref{fig:rof} the redshift of formation as a function of stellar mass for the VIPERS passive red galaxies derived from $D4000_{n}$ (left panel) and  $H\delta_{A}$ (right panel) measurements for a star formation burst of duration $\tau = 0.2$ Gyr and for solar metallicity are shown. The error bars correspond to the different length of the burst 
        ($\tau=$ 0.1, and 0.3 Gyr), 
        
        As shown in Fig. \ref{fig:rof}, the redshift of formation increases with the increasing stellar mass in the same redshift bins. Neglecting the small change in metallicity with stellar mass may somewhat affect the  estimate of $z_{form}$ for high-mass passive red galaxies.  High-mass galaxies may appear older than low-mass galaxies when fitted with a single metallicity model, thus we  assume metallicity dependence on stellar mass. We follow \cite{gallazzi2014} to avoid $z_{form}$
overestimation for high-mass galaxies. We use slightly super-solar metallicities of  0.07 and 0.1 $log(Z/Z_{\odot})$ for $\sim10^{11.1}$, $\sim10^{11.4}$ $\rm{M_{\odot}}$ passive red galaxies, respectively. We decide to ignore the effect of the deviation from solar metallicity at the low-mass end of our sample, which is populated by relatively young galaxies, since for younger stellar population ages the effect of metallicity on both $H\delta_{A}$ and $D4000_{n}$ is negligible (as it can be seen in Fig.~\ref{fig:age}). 
        Since we could exclude that trend would be due entirely to a change of mean metallicity as a function of stellar mass, we conclude that $z_{form}$ increases 
        with increasing stellar mass. 
        This result is independent from the $z_{form}$ calculation method; both  $H\delta_{A}$ and $D4000_{n}$ provide similar trends, which is also a consistency check for our method.

        \begin{table}
                \centering                         
                \begin{tabular}{c| p{1.2cm}| p{1.2cm}| p{1.2cm}| p{1.2cm}}     
                        \hline   
                        \multirow{2}{*}{z} & \multicolumn{2}{c|}{$\sim 10^{10.2}$ $\rm{[M_{\odot}]}$} & \multicolumn{2}{c}{$\sim 10^{11.4}$ $\rm{[M_{\odot}]}$}\\ \cline{2-5}
                        & \multicolumn{1}{c|}{$D4000_{n}$} & \multicolumn{1}{c|}{$H\delta_{A}$}  & $D4000_{n}$ & $H\delta_{A}$   \\
                        \hline 
                        $0.4< z< 0.5$ & $0.79^{+0.06}_{-0.06}$ & $0.88^{+0.05}_{-0.05}$ & - & -\\     
                        $0.5< z< 0.6$ & $0.88^{+0.06}_{-0.07}$ & $0.95^{+0.07}_{-0.07}$ & $1.74^{+0.25}_{-0.19}$ & $2.19^{+0.58}_{-0.40}$\\     
                        $0.6< z< 0.7$ & $1.10^{+0.07}_{-0.08}$ & $1.36^{+0.19}_{-0.12}$ &  $1.54^{+0.10}_{-0.09}$ & $1.78^{+0.34}_{-0.20}$\\  
                        $0.7< z< 0.8$ & $1.16^{+0.10}_{-0.11}$ & $1.16^{+0.11}_{-0.10}$ &  $1.49^{+0.08}_{-0.07}$ & $1.66^{+0.14}_{-0.11}$\\ 
                        $0.8< z< 0.9$ & - & - &  $1.63^{+0.09}_{-0.08}$ & $1.81^{+0.20}_{-0.14}$\\
                        $0.9< z< 1.0$ & - & -  &  $1.68^{+0.09}_{-0.08}$ & $1.81^{+0.20}_{-0.14}$\\
                        
                        \hline\hline                
                        $0.4< z< 1.0$ & $0.98^{+0.07}_{-0.08}$ & $1.08^{+0.10}_{-0.08}$  & $1.62^{+0.12}_{-0.10}$ & $1.85^{+0.29}_{-0.27}$\\  
                        \hline                            
                \end{tabular}
                \caption{Epoch of the last starburst estimated from the comparison of $D4000_{n}$ and $H\delta_{A}$ derived for the VIPERS low-mass and high-mass passive red galaxies with the corresponding values obtained from BC03 model assuming stellar metallicity evolving with mass (0.1 $log(Z/Z_{\odot})$ for the highest stellar mass bin).  The error bars correspond to different burst duration ($\tau = 0.1$ and 0.3 Gyrs).}             
                \label{table:rof}     
        \end{table}
        
        Another interesting  trend is observed in Fig. \ref{fig:rof}: the redshift of formation increases with increasing redshift of observation in the same stellar mass bins. 
        This scaling is especially visible for the less massive systems. This tendency is observed independently of the spectral feature used to derive $z_{form}$, thus it cannot only be explained by a metallicity effect. This result may be a consequence of the progenitor bias; the progenitors of some low-redshift passive red galaxies were still spiral galaxies 
        or merger systems at high redshift and, therefore, were not yet part of the passive sample. Thus, the high-redshift elliptical galaxies would be a biased subpopulation of the low-redshift galaxies, since they include only the oldest progenitors of the low-redshift elliptical galaxies. This bias may result in an underestimation of the true evolutionary track of passive red galaxies. As a consequence, passive red galaxies at higher redshift may have a higher mean $z_{form}$ than a sample observed at lower redshifts~\citep[e.g.,][]{dokkum, moresco}.

        The redshift of formation for VIPERS passive red galaxies in the lowest and highest stellar mass bins derived from the analysis of $D4000_{n}$ and $H\delta_{A}$ are reported in Tab.~\ref{table:rof}.  
        Our estimates imply that massive galaxies ($\rm{log(M_{star}/M_{\odot}) \sim 11.4}$) formed their stars at  $z_{form} \sim 1.7$, while less 
        massive passive red galaxies ($\rm{log(M_{star}/M_{\odot}) \sim 10.2}$) formed their stars at  $z_{form} \sim 1.0$. 
        Taking into account that we are not focused on single galaxies, but we use a stacking to obtain a set of "average" galaxies for different redshifts and stellar mass bins, the estimated redshift of formation should be interpreted as the average property of the bulk of the stellar population of passive red galaxies. The scatter of the $z_{form}$ of stellar populations in individual galaxies is likely distributed around the mean redshift derived for stacked spectra. To quantify the scatter of the $z_{form}$ distribution, we calculated $z_{form}$ for single galaxies on the basis of their $D4000_{n}$ measurements for all stellar mass bins ($10 < \rm{log(M_{star}/M_{\odot})} < 12$) for the redshift bin $0.7 < z < 0.8$. The $\pm 1\sigma$ range for  distribution of $z_{form}$ weighted by $\exp(-\sigma_{z_{form}}^{2}/2)$  is indicated with cyan  areas in  Fig. \ref{fig:rofd4000}. According to this scatter, the derived mean redshift of formation should be interpreted as the representative epoch, while stellar populations in individual passive red galaxies were formed in the range of $- 0.34$, $+ 0.57$ from the mean value obtained for stacked spectra at $0.7 < z < 0.8$.  
        
        The estimated  redshifts of formation for low-mass and high-mass passive red galaxies using the two age spectral indicators (see Tab.~\ref{table:rof}) are in agreement within error bars. Finding similar epochs of formation based on two independent measurements is an important robustness test that reduces the chance of systematic errors.  This also implies that the estimation of one of the age indicators ($D4000_{n}$ or $H\delta_{A}$) is sufficient to establish an epoch of formation for the stellar populations in the intermediate-redshift passive red galaxies.

        \subsection{Comparison with literature}
        
                \begin{figure}[ht!]
                        \centering
                        \centering
                        \includegraphics[width=0.49\textwidth]{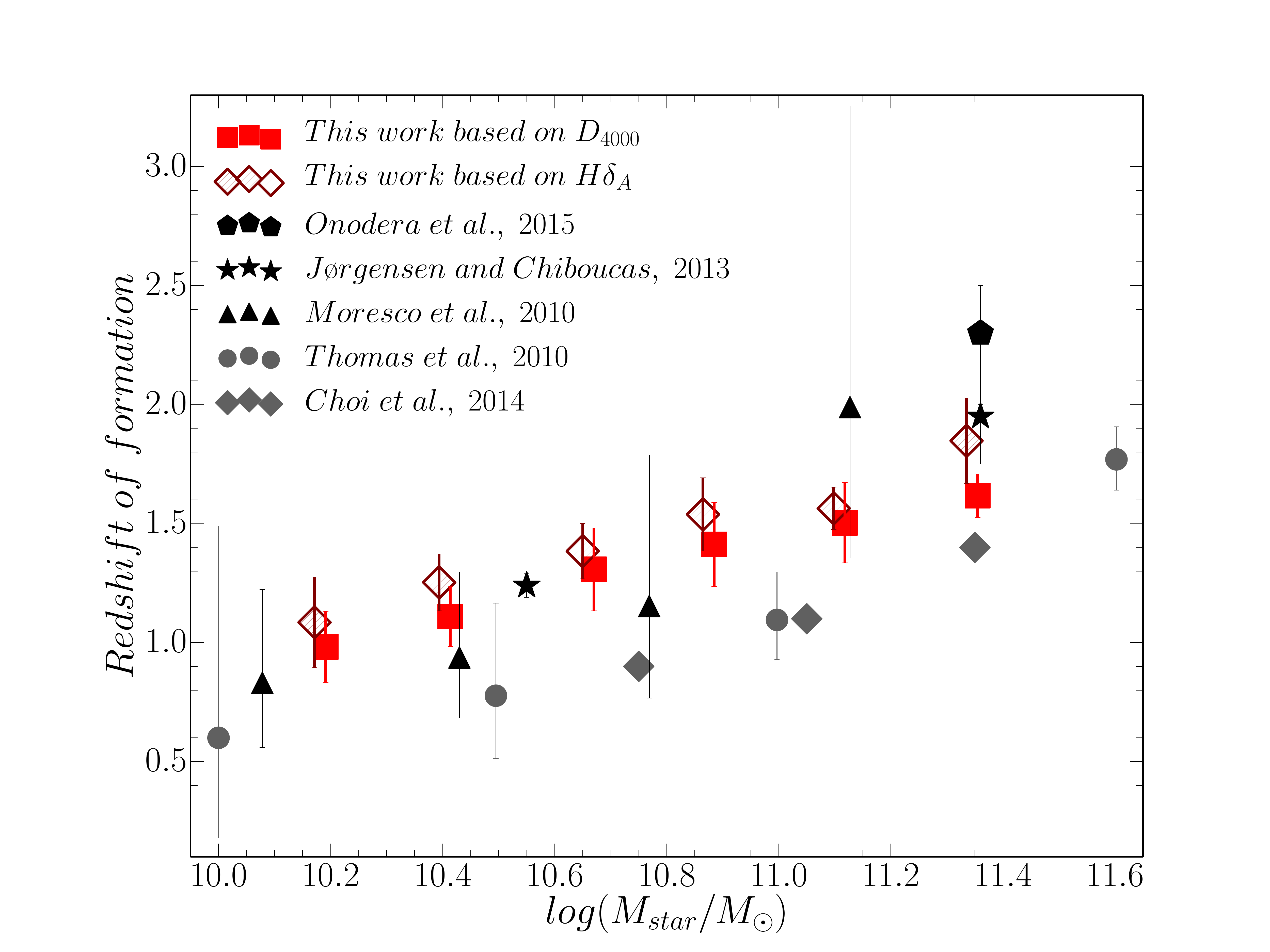}
                        
                        \caption{Mean epoch of the last starburst derived from the  $D4000_{n}$ and $H\delta_{A}$ features estimated for VIPERS passive red galaxies observed at $0.4< z < 1.0$ as a function of stellar mass. Error bars represent the standard deviation within each stellar mass bin. Formation redshifts of stellar populations in intermediate-redshift passive red galaxies derived by~\cite{onodera2015}, \cite{jorg2013}, and \cite{morescoage}  are shown by black pentagon, stars, triangles, respectively. Redshifts of formation at which $50\%$ of the stellar mass of SDSS ETGs was formed as computed by~\cite{thomas2010} are shown with gray circles. Errors correspond to the difference in $z_{form}$ of $50\%$ and $80\%$ of the stellar mass.  Epochs of star formation in local quiescent galaxies established by~\cite{choi} are shown with gray diamonds. }
                        \label{fig:rofcomp}
                \end{figure}
                
        The large majority of studies devoted to obtaining an estimate of the formation epoch for ETGs are focused on nearby and, therefore, relatively bright and easy to observe galaxies.
        However for these objects the fossil record of the past star formation activity is difficult to interpret, mostly because of the severe degeneracy between age and metallicity
        of the stellar population. This degeneracy is expected to become less relevant for high-redshift samples, as the observation epoch gradually approaches the epoch of star formation,
        and therefore a certain number of studies have been carried out recently to  address this question targeting high-z ETGs. The first attempts to derive a formation epoch for ETGs in the redshift range $0.4 < z < 1.3$
        were based on the modeling of the evolution of the fundamental plane relation for cluster ETGs. Assuming that the evolution of the fundamental plane reflects purely the 
        passive evolution of the galaxies mass-to-light ratio, it is possible to estimate the formation epoch for the galaxy population, which is generally located at
        $z_{form} \gtrsim 2$~\cite[e.g.,][]{dokkum2007, saglia, jorg2013}. More recently it has become possible to obtain formation epoch estimates by directly modeling the evolution
        of spectral features in ETGs spectra~\cite[e.g.,][]{morescoage, choi, onodera2015}, but in most cases this exercise has been limited to samples composed of a few tens of galaxies.
        Notwithstanding the large uncertainties associated with these estimates, a general agreement between the various studies exists both on the epoch of ETGs formation and on
        the relative age difference as a function of stellar mass. The overall picture of $z_{form}$ as a function of stellar mass, including our own redshift of formation estimates, which was obtained by averaging together  the estimates for our six redshift bins at each
stellar mass value, is shown in Fig. \ref{fig:rofcomp}.

        Our findings, together with results from previous studies, create a consistent  picture of the $z_{form}$-stellar mass relation.   
        \cite{onodera2015} established $z_{form} $ at the level of 2.3 for a stacked spectrum of 24 massive ($2.3\cdot10^{11}$ $\rm{M_{\odot}}$), quenched galaxies observed in the redshift range $1.25 < z < 2.09$  by comparing its Lick indices with~\cite{thomas2011} models.  
        A similar redshift of formation has been derived for three massive, quenched galaxies in clusters at $z=$ 0.54, 0.83, and 0.89 by~\cite{jorg2013}. Their calculations were also 
        confirmed by an independent analysis on spectral Lick indices. \cite{jorg2013} have found that redshift of formation is higher for higher mass galaxies ($z_{form}\sim1.24$ for galaxies with stellar mass $\sim10^{10.55}$ $\rm{M_{\odot}}$, and $z_{form}\sim1.95$ for galaxies with stellar mass $\sim 10^{11.36}$ $\rm{M_{\odot}}$). These results, 
        obtained at a similar mean redshift as for VIPERS sample, correlates very well with our estimation of redshift of formation. 
        The trend of $z_{form}$ increasing with stellar mass has been also found by~\cite{morescoage}, who have estimated $z_{form} \leq 1$ for intermediate-redshift low-mass ($\rm{log(M_{star}/M_{\odot}) \sim 10.25}$) and  $z_{form} \sim 2$ for high-mass ($\rm{log(M_{star}/M_{\odot}) \sim 11}$) ETGs from zCOSMOS survey.  

        Moreover,~\cite{dokkum2007} have found $z_{form} \sim$ 2.01 after correcting for a progenitor bias for massive ($10^{11}$ $M_{\odot}$) ETGs in clusters observed at $0.18 < z < 1.28$. 
        These results are all consistent with the trend observed for $z_{form}$-stellar mass relation established for the VIPERS passive galaxies.
        
        The epoch of star formation established from the analysis of local passive red galaxies is lower with respect to $z_{form}$ derived from the intermediate-redshift studies at given stellar mass (see Fig. \ref{fig:rofcomp}). \cite{thomas2010} have found that $50\%$ of stellar mass of low-mass  ($\rm{log(M_{star}/M_{\odot}) \sim 10}$) SDSS ETGs has been formed at $z_{form} \leq 1$, while for high mass ($\rm{log(M_{star}/M_{\odot}) \sim 11.6}$) at $z_{form} \sim 2$. 
        This result is in agreement with that obtained by~\cite{choi}, who have found $z_{form} \leq 1.5$ for quiescent SDSS galaxies via a full spectrum fitting of stacked galaxy spectra within a redshift range $0.1 < z < 0.7$. 
        This trend of $z_{form}$ increasing  with stellar mass obtained for local quiescent galaxies is also found for intermediate-redshift galaxies. The $z_{form}$ of SDSS galaxies are somewhat lower at given stellar mass, which may be a consequence of a progenitor bias. 
        
        Our findings confirm that $z_{form}$ of ETGs is shifting to higher values with increasing mass. The transition mass derived for local and intermediate-redshift galaxies~\citep[$3\cdot 10^{10}$ $\rm{M_{\odot}}$ at z $\sim$ 0.2 and $5\cdot 10^{10}$ $\rm{M_{\odot}}$ at z $\sim$ 0.7,][]{kauffmana,pannella} appears to coincide with the 
        region where the $z_{form}$ versus stellar mass relation becomes flatter. Above the transition mass the population of quiescent galaxies is dominated by old, passive red galaxies with 
        no sign of star formation. Thus, for these groups we do not see any major change in their properties and $z_{form}$-stellar mass relation follows a specified trend. On the other 
        hand, below the transition mass, we find passive red galaxies with younger stellar populations. The properties of this population may be affected by the presence of a small fraction 
        of galaxies that only recently became passive red galaxies because below the transition mass the star-forming galaxies represent the dominant population.

        \section{Summary}\label{sec:summary}
        In this work we present the first results of our study of star formation history of passive red galaxies in the redshift range $0.4 < z < 1$.  
        For a reliable estimation of star formation history of passive red galaxies it is essential to face two main challenges: to select a non-contaminated and complete sample of passively evolving galaxies over the wide redshift range and to estimate their stellar ages. To achieve this goal
        \begin{itemize}
                \item We selected a unique, pure sample of 3,991 passive red galaxies in the  redshift range from 0.4 up to 1.0 with stellar masses $\rm{10.00 < log(M_{star}/M_{\odot)} < 12}$ based on a bimodal color criterion with an evolving cut~\citep{fritz} and additional quality ensuring cuts (see Sect. \ref{sec:etg}).
                
                \item We performed a spectral analysis based on $D4000_{n}$ and $H\delta_{A}$ and their dependence on stellar mass and compared them with the data from the local Universe to create a continuous picture of star formation history of passive red galaxies in the redshift range $0.1 < z < 1.0$. 
                
                \item We found that both $D4000_{n}$ and $H\delta_{A}$ display a strong and almost linear dependence on stellar mass (see Fig.~\ref{fig:lick}).

                \item We extended, compared to previous works~\citep{morescoage}, the analysis of the dependence of $D4000_{n}$ on stellar mass and redshift to  higher stellar mass ($\rm{log(M_{star}/M_{\odot}) \sim 11.3}$)  and lower redshift ($z \sim 0$), finding a dependency that is consistent with what is found in the local Universe 
                ($\langle\Delta D4000_{n} \rangle = 0.19 \pm 0.04$, $\langle\Delta H\delta_{A} \rangle = -1.57 \pm 0.44$).

                \item Our analysis confirms the downsizing scenario, as the redshift of formation  increases with stellar mass, and massive galaxies have older stellar populations than less massive galaxies, with metallicity variations with stellar mass providing only a relatively minor perturbation to this overall evolutionary picture. 
                
                \item We found that $z_{form}$ is shifting to higher values with decreasing redshift of observation, which may be a consequence of a progenitor bias. 
                \item We estimated the single-burst star formation epoch as $z_{form} \sim$ 1.7 for massive ETGs (($\rm{log(M_{star}/M_{\odot})} > 11$), while 
                for less massive galaxies we obtained $z_{form} \sim$ 1.0. These results are in agreement with previous estimates based on the modeling of the evolution of the
                fundamental plane relation, or on spectral indicators such as those used in our analysis. We also find a very good agreement between the two estimates of $z_{form}$ 
                obtained on the basis of two independent measurements of age indicators: $H\delta_{A}$ and $D4000_{n}$.

        \end{itemize}

        \begin{acknowledgements}
                The authors want to thank the referee for useful and constructive comments.  We acknowledge the crucial contribution of the ESO staff for the management of service observations. In particular, we are deeply grateful to M. Hilker for his constant help and support of this program. Italian participation in VIPERS has been funded by INAF through PRIN 2008, 2010, and 2014 programs. LG and BRG acknowledge support of the European Research Council through the Darklight ERC Advanced Research Grant (\# 291521). OLF acknowledges support of the European Research Council through the EARLY ERC Advanced Research Grant (\# 268107). AP, KM, and JK have been supported by the National Science Centre (grants UMO-2012/07/B/ST9/04425 and UMO-2013/09/D/ST9/04030). RT acknowledge financial support from the European Research Council under the European Community's Seventh Framework Programme (FP7/2007-2013)/ERC grant agreement n. 202686. EB, FM, and LM acknowledge the support from grants ASI-INAF I/023/12/0 and PRIN MIUR 2010-2011. LM also acknowledges financial support from PRIN INAF 2012. Research conducted within the scope of the HECOLS International Associated Laboratory, supported in part by the Polish NCN grant DEC-2013/08/M/ST9/00664.
        \end{acknowledgements}

        %
        \bibliographystyle{aa} 
        \bibliography{vipers} 
        %
        
        \begin{appendix}
                \section{Estimation of the bias caused by sample selection}
                \label{app:A}
                
\begin{figure}[ht!]
                        \centering
                        \centering
                        \includegraphics[width=0.49\textwidth]{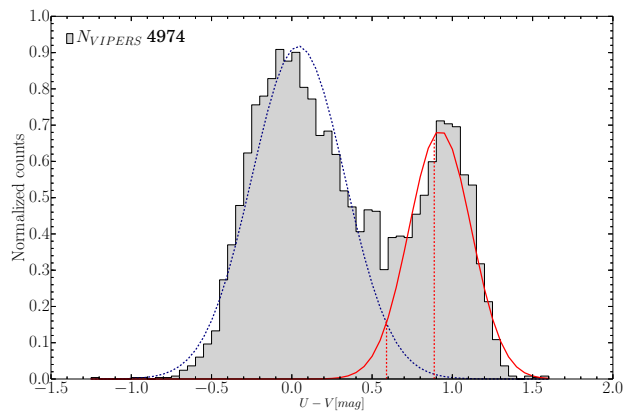}
                        
                        \caption{Distribution of rest-frame $U-V$ color of VIPERS galaxies in redshift bin $0.8 < z < 0.9$ is shown in gray. The dashed blue and solid red lines represent the Gaussian components
                                corresponding to blue and red galaxy populations, respectively.  The  horizontal dashed red lines correspond to the separation of red passive galaxies between the crossing with blue distribution and an evolving cut in $U-V$~\citep{fritz} adopted in this paper. Colors are given in the Vega system. }
                        \label{fig:bimodal}
                \end{figure}

\begin{figure}[ht!]

                        \includegraphics[width=1\linewidth]{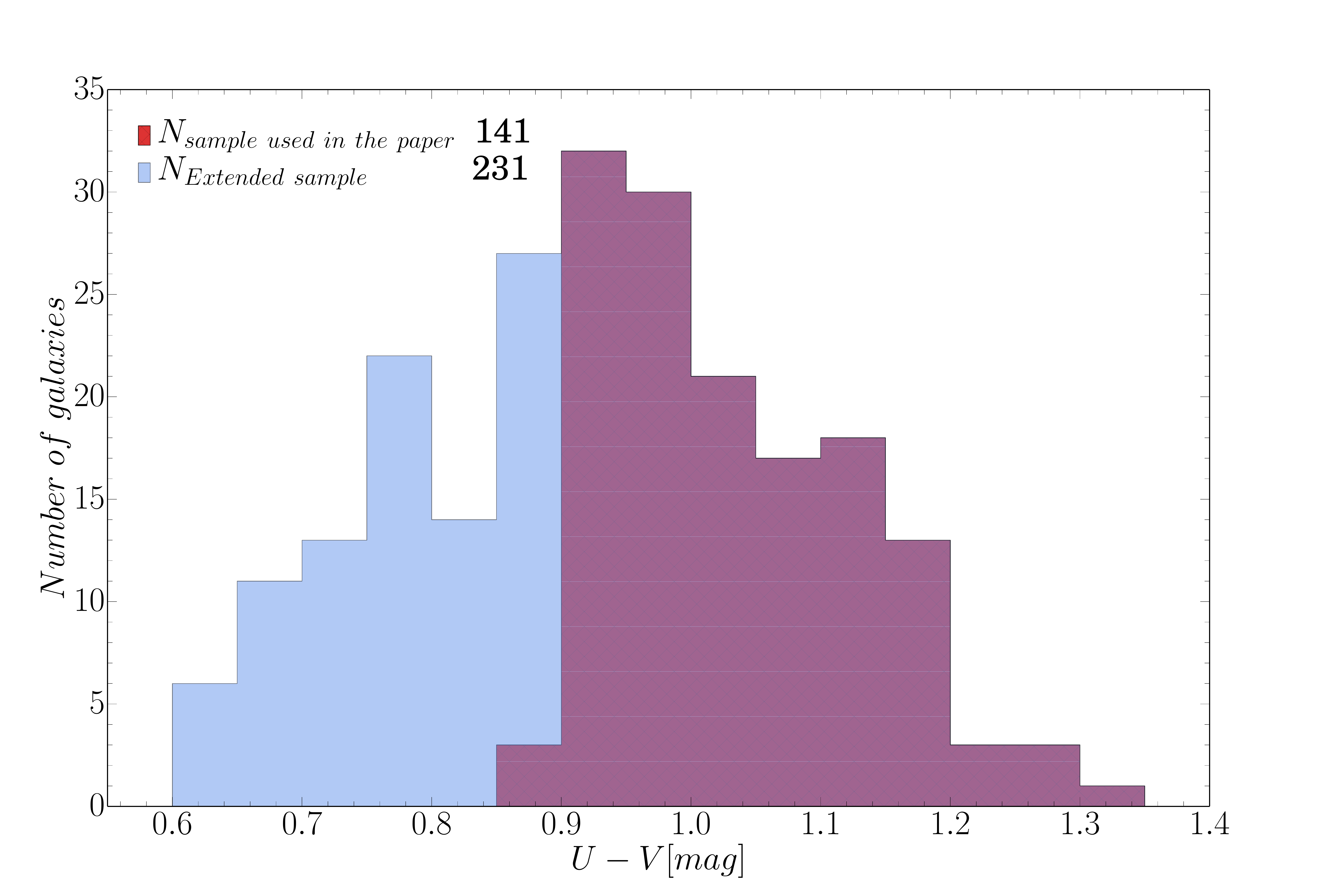}

                        \caption{Distribution of rest-frame $U-V$ color of VIPERS red passive galaxies in stellar mass range $10.50 < \rm{log(M_{star}/M_{\odot})} < 10.75$ and redshift range $0.8 < z < 0.9$.}
                        \label{fig:uv_bimodalmass}
                \end{figure}

                There are a number of different criteria to select passive galaxies. In our paper we used a rather restrictive criterion to minimize the contamination of red passive galaxies by blue, star-forming galaxies. To estimate the possible influence of a sample selection on our results, we extended our group of red passive galaxies by  including less red VIPERS galaxies in the redshift range $0.8 < z < 0.9$.  To select these additional galaxies, 
                we fit the rest-frame $U-V$  color distribution  of VIPERS galaxies in this redshift bin with a two-Gaussian mixture model\footnote{\url{http://scikit-learn.org/stable/modules/mixture.html}} and select a group between the place where the blue and red  distributions  cross and the bimodal cut adopted  in our work (see Fig.~\ref{fig:bimodal}). We selected 13, 90, 44, 26, 8 galaxies fulfilling our additional selection criteria as for original sample (i.e., spectra without any defects in the rest-frame wavelength range $\rm{3850-4250\AA{}}$ and without any sign of star formation activity) for stellar mass bins 10.25 < $\rm{log(M_{star}/M_{\odot})}$ < 10.50, 10.50 < $\rm{log(M_{star}/M_{\odot})}$ < 10.75, 10.75 < $\rm{log(M_{star}/M_{\odot})}$ < 11.00, 11.00 < $\rm{log(M_{star}/M_{\odot})}$ < 11.25, 11.25 < $\rm{log(M_{star}/M_{\odot})}$ < 12.00, respectively. Exemplary distribution of rest-frame $U-V$ color for VIPERS passive galaxies, also including a part of less red population, is shown in Fig.~\ref{fig:uv_bimodalmass}.
                
\begin{figure}[ht!]
                        \includegraphics[width=1\linewidth]{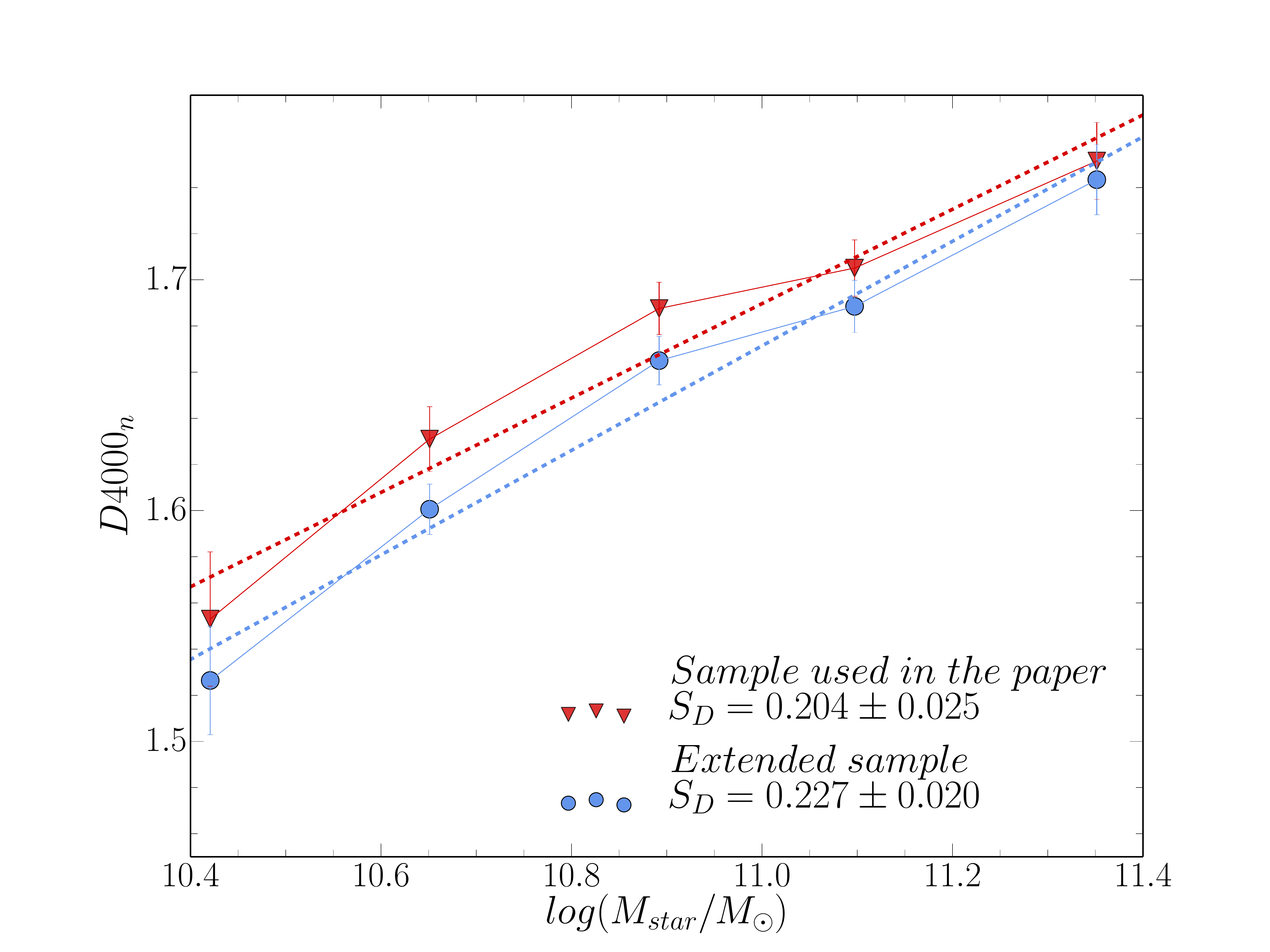}
                        
                        \caption{$D4000_{n}$ as a function of stellar mass for VIPERS stacked spectra of passive red galaxies. The linear fits to the sample used in the paper and to sample including the less red part of the galaxy population are shown as red and blue dashed lines, respectively.  }
                        \label{fig:d4000masstest}
                \end{figure}

                We repeated our analysis; we co-added spectra also including in our sample less red passive galaxies and calculated spectral features in the same way as for the sample used in the paper (see Sect.~\ref{sec:methedology}). The $H\delta_{A}$ obtained for a sample selected in a less restrictive way stay in agreement within error bars with the result obtained for the sample used in the paper for all stellar mass bins. The difference between $D4000_{n}$ is less than $1\sigma$. The $D4000_{n}$ change affects only slightly the $D4000_{n}$-mass relation (see Fig.~\ref{fig:d4000masstest}), resulting in a change in a slope lower than $1\sigma$ ($S_{D}=0.204 \pm 0.025$, $S_{D}=0.227 \pm 0.020$ for the sample used in the paper, and for the extended sample, respectively). Thus, the bias introduced by lower completeness in respect to bluer part of passive galaxies population does not change our results significantly.

                \section{Influence of aging on the slopes of $D4000_{n}$ and $H\delta_{A}$- stellar mass relation        }
                \label{app:B}
        The slopes of relations between two spectral indices used in our analysis ($D4000_{n}$ and $H\delta_{A}$) and stellar mass are similar for passive red galaxies observed in the local Universe and at $z\sim1$. However, if the changes in $D4000_{n}$ (or $H\delta_{A}$) were entirely due to stellar age differences, where high-mass galaxies are $\sim2$ Gyr older than low-mass galaxies, the $D4000_{n}$- stellar mass relation should be steeper at higher redshift, as the $D4000_{n}$- stellar age relation is steeper at lower $D4000_{n}$/ages (see Fig.~\ref{fig:age}). Indeed, the slope of the $D4000_{n}$-stellar age relation is two times higher for VIPERS than for SDSS red passive galaxies ($S_{age}$, Tab.~\ref{table:slopes}). 
                To derive those slopes, we fit the relation in the region of stellar ages obtained on the basis of comparison of $D4000_{n}$ measured for the full VIPERS sample and for SDSS with the values obtained for the grid of synthetic spectra with solar metallicity (see Sect.~\ref{sec:downsizing} and Fig.~\ref{fig:age}). 
                Assuming such a slope and derived age difference, we obtained the slope of $D4000_{n}$-stellar mass relation ($S_{mass}$, Tab.~\ref{table:slopes}). The expected slopes stay in agreement with the slope obtained for $D4000_{n}$-stellar mass relation ($S_{D}$, Tab.~\ref{table:slopes}).  However, if we want to find the same population of red passive VIPERS galaxies in the local Universe, we expect the slope of $D4000_{n}$-stellar mass to  be lower than that observed for SDSS red passive galaxies. This relation is not evolving because the difference in age between low- and high-mass bins changes. This may be an effect of the progenitor bias, which causes the youngest red passive galaxies observed in the local Universe to drop out of the high-redshift sample and, therefore, results in a smaller  observed age difference. The number density of the most massive red passive galaxies is almost constant, while for  low masses it increases by a factor of 2-3 between $z\sim1$ and $z\sim0$~\citep{pozetti,morescocolor}. Thus, the age difference for red passive galaxies at $z\sim1$ is lower than for the local Universe.

                \begin{table}
                        \centering                         
                        \begin{tabular}{| c| p{0.95cm}| p{0.7cm}| p{0.95cm}| p{0.7cm}| p{0.95cm}| p{0.7cm}| }     
                                \hline   
                                z & \multicolumn{2}{|c|}{$S_{age}$ $\pm$ $\sigma_{S_{age}}$} & \multicolumn{2}{|c|}{$S_{mass}$  $\pm$ $\sigma_{S_{mass}}$} & \multicolumn{2}{|c|}{$S_{D}$ $\pm$ $\sigma_{S_{D}}$} \\
                                \hline 
                                \multicolumn{7}{|c|}{$D4000_{n}$}   \\
                                \hline 
                                $z\sim 0.15$ & \multicolumn{2}{|c|}{0.066 $\pm$ 0.004} & \multicolumn{2}{|c|}{0.151  $\pm$ 0.014} &  \multicolumn{2}{|c|}{0.141 $\pm$ 0.002} \\ 
                                \hline                                
                                $z\sim0.75$ &  \multicolumn{2}{|c|}{0.136 $\pm$ 0.008} &  \multicolumn{2}{|c|}{0.172 $\pm$ 0.010} &  \multicolumn{2}{|c|}{0.164 $\pm$ 0.031} \\     
                                \hline       
                                \multicolumn{7}{|c|}{$H\delta_{A}$}   \\
                                \hline 
                                $z\sim0.15$ &  \multicolumn{2}{|c|}{-0.329 $\pm$ 0.010} &  \multicolumn{2}{|c|}{-1.119 $\pm$ 0.034} &  \multicolumn{2}{|c|}{-1.141 $\pm$ 0.023} \\ 
                                \hline                                
                                $z\sim0.75$ &  \multicolumn{2}{|c|}{-1.480 $\pm$ 0.116} & \multicolumn{2}{|c|}{ -1.399 $\pm$ 0.110} &  \multicolumn{2}{|c|}{-1.353 $\pm$ 0.376} \\     
                                \hline                           
                                \end{tabular}
                                \caption{Slopes of $D4000_{n}$-stellar mass relation ($S_{D}$) are in agreement with those expected from the $D4000_{n}$-stellar age relation ($S_{mass}$). The sames goes for $H\delta_{A}$.  }             
                                \label{table:slopes}     
                                \end{table}

        \end{appendix}
\end{document}